\documentclass[reprint,amsmath,amssymb,aps,prapplied,floatfix]{revtex4-2}
\makeatletter
\usepackage[hidelinks,colorlinks=true,linkcolor=blue,citecolor=blue, urlcolor = blue]{hyperref}
\usepackage{bookmark}
\newcommand*{\rom}[1]{\expandafter\@slowromancap\romannumeral #1@}
\usepackage{parskip}
\usepackage{graphicx}
\usepackage{dcolumn}
\usepackage{bm}
\usepackage{comment}
\usepackage[normalem]{ulem}
\usepackage[utf8]{inputenc}
\usepackage{hyperref}
\usepackage{lineno}
\usepackage{changepage}
\usepackage{multirow}
\usepackage{array}
\usepackage{float}
\usepackage{amsmath}
\usepackage{titlesec}
\usepackage{cleveref}
\usepackage{feynmp-auto}
\usepackage{pgfplots}
\pgfplotsset{compat=1.15}
\usepackage{mathrsfs}
\usetikzlibrary{arrows}
\begin{document}
\preprint{APS/123-QED}
\title{\texorpdfstring{Search for nonextensivity in electron-proton interactions at $\sqrt{s}$ = 300 GeV}{TEXT}}
\author {Soumya Sarkar}
\email{soumya.sarkar@students.iiserpune.ac.in}
\affiliation{Indian Institute of Science Education and Research, Pune, India}
\author {R. Aggarwal}
\email{ritu.aggarwal1@gmail.com}
\affiliation{USAR, Guru Gobind Singh Indraprastha University, East Delhi Campus, 110092, India}
\author {M. Kaur}
\email{manjit@pu.ac.in}
\affiliation{Department of Physics, Panjab University, Chandigarh 160014, India\\
Department of Physics, Amity University, Punjab, Mohali {140306}, India}
\date{\today}
\frenchspacing
\begin{abstract}
Study of canonical entropy in electron-proton interactions at $\sqrt{s}$ = 300 GeV is presented. The precision data collected by the H1 experiment at the HERA in different ranges of invariant hadronic mass $W$ and the squared four-momentum exchange $Q^{2}$ in $ep$ interactions have been analyzed in the ensemble theory approach. The canonical partition function relates to the multiplicity distribution which is often studied in collider experiments. We use the canonical ensemble partition function to explore the dynamics of hadron production in $ep$ interactions by devising different methods to find the entropic parameter and the collision temperature. The inverse slope of the transverse momentum spectrum of produced hadrons also relates to the temperature. In the recent past, the CMS, ATLAS and ALICE experiments at the LHC have studied the charged hadron transverse momentum and particle distributions in proton-proton and proton-nucleus interactions by using the Tsallis function within this approach. A detailed investigation into the role of the system volume and relation amongst different dynamical parameters reveals interesting results.
\end{abstract}
\maketitle
\section{\label{sec:level1}INTRODUCTION}
The application of ensemble theory of statistical mechanics to particle interactions gives interesting results to understand the interaction dynamics. High energy collisions of subatomic particles such as pions, protons, electrons and the heavy ions etc. produced at the particle accelerators are studied systematically with high precision.~The production of particles, both known and unknown in such collisions are also studied in terms of physics principles emerging from the ensemble theory. Many of the physical observables such as hadron multiplicity, rapidity and transverse momentum etc., which are  few of the different outcomes from these collisions are understood by modeling through the concepts of thermodynamics and statistical mechanics \cite{kanki1989,kittel2005soft,TSALLIS1998,DREMIN2001,ellis1985,Gio2005}. Exciting new results from the estimated behavior of these observables as a function of center-of-mass energy have emanated from the detailed investigations allowing the predictions for the future experiments in pursuit of discovery of new particles \cite{Gio2005,Bar2011,Marquis2015}. Ensemble theory has been used in several studies of particle production. The multiplicity of particles produced in a collision and its relation with the thermodynamic temperature of a collision are used to access dynamics of the particle interactions. Visualising a particle interaction as a micro-canonical or canonical or grand-canonical ensemble is one of the interpretations which has led to the understanding of particle-production mechanisms. For example the canonical partition function relates to the multiplicity distribution often measured in collider experiments and the transverse  momentum spectrum reveals information on the early thermal or close to thermal properties of the hot state of such collisions \cite{kk,Shen2019,Wibig_2010,Marquis2015} etc. 
\\
Standard statistical mechanics of Boltzmann-Gibbs which treats entropy as an extensive property of the system and the models based on this have not been very satisfactory. The possible sources of deviations included intrinsic and non-statistical effects, in particular fluctuations in the properties of entropy. A genealized form of the entropy was postulated by C.Tsallis introducing a redefinition \cite{TS1} as follows;
\\
The standard expression for entropy is $S = -k \sum_{i=1}^{W} p_{i} \ln p_{i}$ where $W\in N$ is the total number of possible configurations corresponding to \{$p_{i}$\} associated probabilities and $\sum_{i=1}^{W} p_{i}$ =1. The entropy $S$ is thus an extensive property of any thermodynamical system. However, in the redefinition of entropy, $S_{q} = \frac{k}{q-1}\sum_{i=1}^{W} p_{i} (1-p_{i}^{q-1})$, where  $k$ is a positive constant, $q>1$ makes the entropy a nonextensive property of the system. The  $q$, also known as the entropic index, characterizes the degree of nonextensivity and the additive entropy rule stands modified;
\begin{equation}
S_{q}(A+B) = [S_{q}(A)] + [S_{q}(B)] + (q-1)[S_{q}(A)][S_{q}(B)]
\end{equation}
where $A$ and $B$ are two independent systems and $q >$ 1 is a measure of the nonextensivity of entropy. For $q$=1, one recovers back the  Boltzmann-Gibbs(BG) statistics.

Thermostatistics has been applied to describe the particle production in high-energy particle interactions. Predictions of a thermodynamical model of hadron production for multiplicity distributions in $e^{+}e^{-}$ annihilation at LEP and PEP-PETRA energies have been used to establish a two-step process resulting into the clan structure \cite{becattini1995,becattini1996,2000llvh.book.....G,Gio2005}. The clan model was introduced in order to interpret the wide occurrence of the negative binomial regularity of charged particle multiplicity distribution in high energy reactions \cite{gio1986NB,Gio2005}. In a reaction any produced particle originates from a primary particle (or a parton), named ancestor. All the particles with a common ancestor form a clan. The clans have no mutual interaction. The reaction dynamics may be such that all the produced particles may be correlated.
Thermostatistical aspects play an important role in the investigations in high-energy collisions of two particles or heavy-ions to predict and understand the form of statistical equilibrium. One of the signatures of “thermal” multi-particle production is the exponential form of the transverse energy distribution of the produced hadrons. The slope parameter of this distribution can be interpreted in terms of a temperature of the final state \cite{Marquis2015,Barnaföldi_2015,Rui-2018}.

The slope of this distribution is interpreted in terms of a temperature of the final state. The particle multiplicities and transverse energy distribution are found to be consistent with such an interpretation, at lower energies. At higher energies, the thermal interpretation of the transverse energy spectra is modified and described well by the nonextensive thermostatistics of Tsallis \cite{TS1,BEDIAGA2000,BECK2000}. At energies ($\sqrt{s}$ =  200 GeV and beyond) in $pp$ collisions a significant deviation from the exponential transverse energy distribution, together with the violation of KNO \cite{kno} scaling law was encountered.

The nonextensive generalization of the statistical mechanics has gained importance in describing the collisions at collider experiments. It has been extensively studied in different type of high-energy collisions \cite{Pereira2007,PEREIRA2009,CONROY2010,Cleyman2012,Gervino2012,Santos2014} up to the highest energy data from $e^{+}e^{-}$ collisions, $pp$ collisions and for heavy-ion collisions \cite{Wibig_2010,khachatryan2010transverse,Aad2011,DENTERRIA2011,CLEYMANS2013351}. Most of these studies have focused on the particle properties and in particular the charged particle multiplicities and the transverse momentum spectra \cite{Marquis2015,Cleymans2017kzp,DEPPMAN2021,Tao2022}. In the recent past, the CMS, ATLAS and ALICE experiments at the Large Hadron Collider(LHC) have studied the charged hadron transverse momentum distributions in proton-proton and proton-nucleus interactions by using the Tsallis function \cite{TS1,s39,s40,alice2011production}. In addition to the description of the transverse energy spectra, the extended statistical approach to describe multiparticle production can be used to predict the temperature.
 
In this paper we present first study in terms of canonical entropy from the multiplicities in different kinematic regions of electron-proton collisions at the Hadron-Electron-Ring Accelerator (HERA). The precision data of proton-proton interactions at the Intersecting Storage Rings (ISR) at different energies have been analyzed for the validation. We use these two different methods to describe the properties of multiplicity distributions and their temperature dependence. The section \ref{sec:level2} describes the concepts, the equations used for computing entropic index and the temperature with some inputs from \cite{AGUIAR2003}. Details of the data being analyzed are given in section \ref{sec:level3}. Methodology developed for determining the parameters are give in \ref{sec:level4}. Results and discussion are presented in \ref{sec:level5}, followed by conclusion in section \ref{sec:level6}.

\section{\label{sec:level2} MULTIPLICITY AND GENERALISED CANONICAL PROBABILITY}
An important observable in the particle interactions is the multiplicity distribution of the produced hadrons. One of the most popular ways of expressing the probability distribution of these hadrons is the negative binomial distribution \cite{nbd1986,VanHove:1986xr}, which is known to arise with some specific processes such as accompanying the Bose-Einstein particles with different sources. Following subsection outlines the distribution. 

\subsection{Negative binomial distribution}
The multiplicity of particles produced in an interaction can be described in terms of the negative binomial distribution(NBD), which is the following probability law for multiplicity $n\ge0$;
\begin{equation}
P_n = \frac{k(k+1).....(k+n-1)}{n!}\left(\frac{\langle{n}\rangle}{\langle{n}\rangle+k}\right)^n \left(\frac{k}{\langle{n}\rangle+k}\right)^k\label{eq:nbd}
\end{equation} 
where ${\langle{n}\rangle}$ is the mean number of produced particles called the average multiplicity and $k>0$ is related to the variance $D$ of the distribution as:
\begin{equation}
\frac{1}{k}=\frac{D^2-\langle{n}\rangle}{\langle{n}\rangle}^{2} \label{eq:k_NBD}
\end{equation}
The negative binomial distribution has been extensively used to fit the multiplicity distributions by nearly all high energy physics experiment. It remarkably well described the data at lower energies, but showed significant deviations at higher energies. The weighted combinations of NBDs have been used to improve its applicability \cite{ALNER1986,DERRICK1986,UA51987}.
\subsection{\texorpdfstring{The $q-$}{TEXT}statistics}
It is well known that several features of particle production in high energy hadronic collisions can be described by thermostatistical models. A phase of nuclear matter consisting of quarks and gluons in such collisions is expected to lead to some form of statistical equilibrium, subsequently leading to the formation of locally thermalised source of particles. The generalised statistical mechanics (with $q >$1) affects the hadronic multiplicity distribution in such collisions.
In usual thermostatistics the generalized entropy $S$ is defined \cite{AGUIAR2003} by introducing the $q$-entropy;
\begin{eqnarray}
    S = \frac{1-\sum_{a}P_{a}^q}{q-1} \label{eq:ent} \\ 
    \sum P_{a} = 1
\end{eqnarray}
where $P_{a}$ is the probability of microstate $a$. Total probability being normalized to unity. Any physical observable ${O}$ in this approach has a $q$-biased average value defined as;
\begin{equation}
\langle{O}\rangle = \frac{\sum_{a}O_{a}P_{a}^q}{\sum_{a}P_{a}^q}
\end{equation}
And the $q$-biased microstate probability;
\begin{equation}
\tilde{P}_{a} =  \frac{P_{a}^q}{\sum_{a}P_{a}^q}
\end{equation}
is the probability to be used in calculation of physical quantities. In the limit when the entropic index $q\rightarrow1$ the normal Boltzmann-Gibbs-Shannon entropy is recovered \cite{Ramshaw_2020}. The equilibrium probabilities $P_{a}$ are determined by maximising the entropy under the requisite constraints on the charge and energy conservation. For a fixed value of energy $E$, the conserved electric charge is $Q$. The constraints are implemented through the variational principle, using the method of undetermined Lagrange multipliers and using the normalisation conditions on the energy ${E_a}$ and charge ${Q_a}$ of a microstate $a$. The variational principle gives;
\begin{equation}
    \delta S+ c_{1}\sum_a\delta p_a-c_2\delta E + c_{3}\delta Q = 0,
\end{equation}
where
\begin{eqnarray}
    E = \sum_a E_a\tilde{p}_{a}, \\
    Q = \sum_a Q_a\tilde{p}_{a} 
\end{eqnarray}
 The constants $c_{1}$, $c_{2}$ and $c_{3}$ are Lagrange multipliers. Temperature T and chemical potential are related to $c_{2}$ and $c_{3}$ as:
\begin{eqnarray}
    c_{2} = \frac{1}{T} = \left(\frac{\partial S}{\partial E}\right)_{Q,V} \\
    c_{3} = \frac{\mu}{T} = -\left(\frac{\partial S}{\partial Q}\right)_{E,V}
\end{eqnarray}
The $\tilde{P}_{a}$ distribution can be obtained by solving the variational principle and is given below;
\begin{equation}
\tilde{P}_{a} =  \frac{({\exp_{q}[-\beta(E_{a}-\mu Q_{a})]})^q}{\sum_{a} \left({\exp_{q}[-\beta(E_{a}-\mu Q_{a})]}\right)^q} \label{eq:tilP}
\end{equation}
The denominator in Eq. (\ref{eq:tilP})
is called as the generalised partition function, $Z_{q}(\beta, \mu, V)$. The $q$-potential function, $\exp_{q}$ is defined by
\begin{equation}
 \exp_{q}(A) = [1-(q-1)A]^{-1/(q-1)} \label{eq:exp1}  
 \end{equation}
 $\beta$ is related to the temperature $T$ by;
\begin{equation}
  T = \frac{\beta^{-1} + (q-1)(E-\mu Q)}{1+(1-q)S} \label{eq:T}   
\end{equation}
and when $q\sim1$;
 \begin{equation}
 [\exp_{q}(A)]^q \sim \exp[A + (q-1)(A+A^{2}/2)] \label{eq:exp2}  
 \end{equation}
The Lagrange multipliers express the temperature and chemical potential. A characterisation of thermal equilibrium connects the so called 'physical' temperature via
\begin{equation}
\tilde{T} = \beta^{-1} + (q-1)(E-\mu Q) \label{eq:tilT}
\end{equation}
Chemical potential $\mu$ controls the average charge $Q$ in the grand canonical approach discussed. For small values of $Q (=1)$, when fluctuations about the mean become significant, the canonical treatment is preferred and $\mu$=0. The generalised partition function for fixed charge becomes; 
\begin{equation}
Z_{q}(\beta,Q,V) = \sum_a \delta(Q-Q_a)\left({\exp_{q}[-\beta(E_{a})]}\right)^q
\end{equation}
where $\delta(Q-Q_a)$ is the Kronecker delta. The generalised canonical probability is given as;
\begin{equation}
\tilde{P}_{a} = \frac{\delta(Q-Q_{a})}{Z_{q}(\beta, Q, V)}\left[{\exp_{q}(-\beta(E_{a})}\right]^q 
\end{equation}
The probability that a system has exactly $N$ number of particles in the Tsallis statistics, with $N_a$ particles in a given state $a$ is;
\begin{equation}
P_N = \sum_a \delta(N-N_a)\tilde{P}_{a}
\end{equation}
The $N$ particle partition function can then be written as;
\begin{equation}
Z_{q}^{(N)}(\beta,\mu,V) = \sum_a\delta(N-N_a)\left[exp_{q}(-\beta(E_{a}-\mu Q_{a}))\right]^q
\end{equation}
 where
\begin{equation}
P_{N}=\frac{Z_q^{(N)}}{Z_q}
\end{equation}
The generating function for the multiplicity distribution can now be defined by;
\begin{equation}
  F(t)\equiv\sum_{N=0}^{\infty}t^{N}P_{N}
\end{equation}
The multiplicity distribution for $N$ particle system can then derived from the generating function. For example, for the negative binomial distribution 
the generating function is;
\begin{equation}
F_{NB}(t) = \left[1-\frac{\langle N\rangle}{k}(t-1) \right]^{-k} = \exp_{q}[\langle N\rangle (t-1)]\label{eq:FNB}
\end{equation}
With $q=1+1/k$, the Eq.(\ref{eq:FNB}) gives the negative binomial distribution. For large values of $k$ or small values of $q$ we have the asymptomatic behavior of $F_{NB}(t)$ given by:
\begin{equation}
F_{NB}(t) \approx exp{[\langle N\rangle(t-1)+\frac{\langle N\rangle^2}{2k}(t-1)^2]} \label{eq:FNBapprox}
\end{equation}

When these results are applied to an ideal relativistic gas to study the Tsallis statistics of hadron gas, it is observed that with $q>1$ the generalized partition function $Z_q$ becomes infinite. Thus there is no ideal Tsallis gas with $q>1$ and the $q$-statistics of an ideal relativistic gas cannot be defined. However, the case $q>1$ is the focus of interest in high energy collisions. For an ensemble of $N$ particles, the generalized ideal gas partition function $Z_q^{(N)}$ has a well-defined integral representation in terms of the corresponding Boltzmann–Gibbs function subject to the condition that $N<\frac{1}{3(q-1)}$. Particle production above this limit is the cause of divergence in the partition function of the Tsallis ideal gas. One way to include such interactions is to introduce a “Van der Waals” excluded volume simulating the effect of hard-core potentials among particles. It tries to model the effect of non-zero volume of particles of the system as opposed to ideal gas where the volume of particles are taken to be zero \cite{yen1997,braun1999}. Also the ideal gas system is non-interacting but Van der Waals gas takes into account the interaction by considering the particles as hard-spheres. The repulsive interactions among the produced particles are thus responsible for the divergence in partition function of the Tsallis ideal gas. To include such interactions, the Van der Waals excluded volume was introduced. This Van der Waals gas model has been used to study particle abundances in high energy heavy ion collisions as it gives an appropriate description of the particle dynamics in the gas. Van der Waals gas model as well as the Boltzmann-Gibbs statistics with extensive entropy ($q$=1) produce a multiplicity distribution with narrower width as compared with the experimental distribution. However, with Tsallis statistics, increasing $q$ very slightly ($q>1$) a much broader distribution is obtained and in good agreement with the experimental data. This result is true for data at different energies.

\subsection{Excluded volume for N-particle system}
\begin{figure}
\centering
\includegraphics[scale=0.35]{ 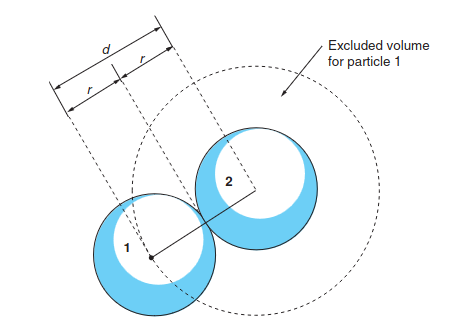}
\caption{Schematic figure of excluded volume with two hard spheres \cite{mortimer2000physical}.}
\label{fig:ExVol}
\end{figure}

The ideal gas description proved to be inadequate in many ways. A primary reason being it does not consider the interaction and finite volume of its constituent particles. The simplest way to add an interaction and also a finite volume is to consider the repulsive hard-core or hard-sphere potential. In this case, the volume which remains inaccessible to other particles of the system as a result of the presence of the first particle, is called the excluded volume. For example, the excluded volume of a single hard sphere (particle) is eight times its volume. If the system has two identical hard spheres then the excluded volume gets distributed among the two \cite{mortimer2000physical}. Assuming that the two spheres are identical, it becomes four times the volume of each sphere, as shown in the schematic Fig. \ref{fig:ExVol}. For a hard-core sphere the excluded volume depends on the radius of the spheres. In principle all particles can have their own hard-core radii, subject to the restrictions that; \\
i) $V_{ex} << V$, with V is the total volume of the system and $V_{ex}$ the total excluded volume of all particles. \\
ii) Also as described in Ref. \cite{oliinychenko2012investigation}, the radii should not be too small to circumvent any contradiction with Lattice Quantum Chromodynamics (LQCD). 
This results in having certain flexibility in defining the value of excluded volume or hard core radii as long as these conditions are satisfied. It has been observed that excluded volume remains nearly constant for nucleon-nucleon scattering. For nucleon-nucleon scattering this value of hard-core radius ($r_0$) is observed to be around $r_0=0.3~fm$ \cite{oliinychenko2012investigation}.
For the current analysis we take the value of excluded volume, $v_0$=0.368 fm$^3$, same as used in \cite{AGUIAR2003} for all the analysis of $pp$ interactions. We use the same value for analysis of $ep$ interactions as well since for all-hadron final state a constant value of $v_0$ can be used, as mentioned earlier. To further understand the effect of Tsallis and Van der Waals corrections to the ideal gas, necessary details as described by Aguiar and Kodama in Ref. \cite{AGUIAR2003} are reproduced below:

The starting point is the relativistic ideal gas. For a relativistic ideal gas the Boltzmann-Gibbs grand canonical partition function is given by:
\begin{equation}
    Z(\beta,\mu,V) = exp\left(V\sum_{i=1}^{h} \Phi_i(\beta)exp(\beta\mu q_i)\right), \label{eq:relpartfunc}
\end{equation}
where
\begin{equation}
    \Phi_i(\beta) = \frac{g_i}{2\pi^2}\frac{m_i^2}{\beta}K_2(\beta m_i), \label{eq:phi}
\end{equation}
$g_i$ denotes the degeneracy factor of $i$-th hadron species with mass $m_i$ and charge $q_i$. $h$ denotes the total number of hadrons. $K_2(z)$ is the modified Bessel function of second kind. $\beta$ denotes the inverse temperature, $\mu$ denotes the chemical potential and $V$ the volume of the system. Hence the $N$-particle partition function is given by:
\begin{equation}
    Z^{(N)}(\beta,\mu,V) = \frac{1}{N!}\left (V\sum_{i=1}^h \Phi_i(\beta)exp(\beta\mu q_i)\right)^N \label{eq:relpartfuncN}
\end{equation}
To have Van der Waals-type effects in relativistic gas an excluded volume $v_0$ is introduced, which changes $V$ to $V-Nv_0$ to take into account the finite volume of particles. The partition function got modified as:
\begin{equation}
    Z^{(N)}(\beta,\mu,V)\rightarrow Z^{(N)}(\beta,\mu,V-Nv_0)\Theta(V-Nv_0)
\end{equation}
The Heaviside $\Theta$ function limits the number of particles to be fitted in a particular volume V i.e $N<V/v_0$. The partition function of a Relativistic Van der Waal's gas is given by:
\begin{equation}
    Z(\beta,\mu,V) = \sum_N \frac{1}{N!}n(\beta,\mu)^N (V-Nv_0)^N \Theta (V-Nv_0)
\end{equation}
Here $n(\beta,\mu)$ denotes the particle density and its expression can be derived from ideal relativistic gas:
\begin{equation}
    n(\beta,\mu) = \sum_{i=1}^h \Phi_i(\beta)exp(\beta\mu q_i) \label{eq:n}
\end{equation}
In the large $V$ asymptotic limit, partition function for relativistic ideal gas reduces to:
\begin{equation}
    Z(\beta,\mu,V) = exp\left\{\frac{V}{v_0}W[v_0n(\beta,\mu)]\right\} \label{eq:vanpartfunc}
\end{equation}
$W(x)$ is the Lambert function. Hence from Eq. (\ref{eq:vanpartfunc}) the q-exponential partition function and hence generating function can be derived as:
\begin{equation}
        Z_q(\beta,\mu,V) = \int_0^\infty dxG(x)exp\left\{\frac{V}{v_0}W[v_0n(x\beta,\mu)]\right\} 
\end{equation}
\begin{equation}
     F(t) = \frac{1}{Z_q(\beta,\mu,V)} \int_0^\infty dxG(x) exp\left\{\frac{V}{v_0}W[tv_0n(x\beta,\mu)]\right\}
\end{equation}
 The partition function $Z_{q}$ then remains convergent as long as $q<1+\frac{v_{0}}{3V}$. To understand the effect of  Tsallis statistics on the multiplicity distribution, the case in which $q-1$ and $v_0$ are both small, the generating function for the Van der Waals–Tsallis relativistic gas can be written as:
\begin{equation}
\begin{split}
F(t) \approx exp\{(t-1)Vn\left[(1+(q-1)\xi(Vn\xi-1)-2v_0n\right] \\ + (t-1)^2(Vn)^2\left[(q-1)\frac{\xi^2}{2}-v_0/V\right]\}\label{eq:F}
\end{split} 
\end{equation}
where $\xi$ in Eq.(\ref{eq:F}) is given by;
\begin{equation}
\xi = -\frac{\beta}{n}\frac{\partial n}{\partial \beta} \label{eq:xi} 
\end{equation}
 and $n\equiv n(\beta,\mu)$ is given by Eq.(\ref{eq:n}).On comparing Eq.(\ref{eq:F}) with generating function for negative binomial distribution, equation(\ref{eq:FNBapprox}), the values of $\langle N \rangle$ and $k$ can be deduced and is given by:
\begin{eqnarray}
\langle N\rangle &=& n_0+(q-1)n_0\xi(n_0\xi-1)-\frac{2v_0}{V}n_0^2 \label{eq:Nbar}\\
\frac{1}{k} &=& (q-1)\xi^2-\frac{2v_0}{V} \label{eq:k}
\end{eqnarray}
Where $n_0 \equiv Vn$ denotes the number of particles at a fixed temperature ($n$ as defined by Eq.(\ref{eq:n}) and $\xi \equiv \xi(\beta,\mu) = -\frac{\beta}{n_0}\frac{\partial n_0}{\partial \beta}$ as given by Eq.(\ref{eq:xi}). For further analysis, we take $\mu=0$. Hence, $n_0$ is given by:
\begin{equation}
    n_0 = V\sum_{i=1}^h \Phi_i(\beta) \label{eq:n0}
\end{equation}
All the equations have been adopted from \cite{AGUIAR2003}. Using Eqs. (\ref{eq:Nbar},\ref{eq:k}) the temperature, $\beta^{-1}$ and non-extensivity parameter, $q$ can be derived, from experimental quantities namely average charged multiplicity, $\langle N \rangle$ and $k$, which are related to width and shape of the NBD distribution, as given in Eq. (\ref{eq:k_NBD}).

\section{\label{sec:level3} DATA USED} 
The primary aim of the analysis is to study the hadronic multiplicities in $ep$ interactions at HERA energy using the canonical partition function with the q-statistics. This is the first analysis with this data. The analysis of data from $pp$ interactions at ISR energies also has been done for the same reason, in addition to it being used for the purpose of validation. Beyond these energies, the KNO scaling \cite{kno} violations were reported as mentioned in Ref. \cite{UA51988}. A separate analysis of the $pp$ data for $\sqrt{s} >$ 200 GeV is under consideration. 
\vspace{-0.6cm}
\subsection{Electron-proton collisions at HERA}
The data used for the present analysis were collected with the H1 detector \cite{H1D} at the HERA storage ring at DESY during the 1994 running period which recorded collisions of positrons, with an incident energy of 27.5 GeV, and protons with an energy of 820 GeV giving $\sqrt{s}$=300 GeV. Multiplicity distributions measured in four different kinematic regions in $W$= 80-115, 115-150, 150-185, 185-220 GeV for charged particles and in four different pseudorapidity sectors; 1$<\eta^{*}<\eta_{c}^{*}$ with $\eta_{c}^{*}$ = 2,3,4,5 have been analyzed.
\vspace{-0.7cm}
\subsubsection{\bf{Kinematics of deep inelastic scattering}} 
\begin{figure}
\centering
\includegraphics[scale=0.35]{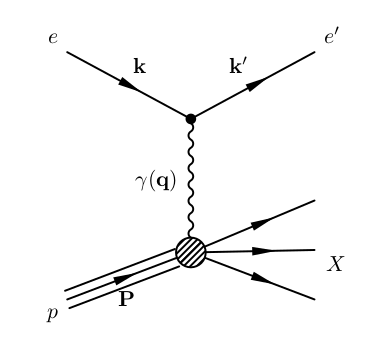}
\caption{Deep Inelastic Scattering of lepton on hadron.}
\label{fig:Dis}
\end{figure}
In deep inelastic scattering, a high-energy lepton  scatters off a hadron after interaction with one of its constituents through a virtual photon or a weak boson as shown in Fig. \ref{fig:Dis}. Let the initial 4-momentum of lepton be $k$ and final 4-momentum $k'$. Initial 4-momentum of proton be $p$, fraction of proton momentum carried by the struck quark as x and final 4-momentum of the hadronic system be $p'$. Following invariant variables can be defined 
\begin{gather}
s=(p+k)^2 \\
t = (p-p')^2 \\
Q^2 = -q_\gamma^2 = -(k-k')^2 \label{QSq}\\
y = \frac{p.q_\gamma}{p.k'}  \label{Y}\\
W^2 = (p')^2 = (p+q_\gamma)^2\
\end{gather}
Where $s$ is the center-of-mass energy squared, $t$ is the four-momentum transferred squared between proton and final state hadronic system, $Q^2$ is negative square of the four-momentum transferred ($q_\gamma$) from the electron to the proton, $y$ is the inelasticity of the scattered lepton and $W^2$ is the invariant squared mass of final state hadrons. The energy-momentum conservation demands that:
\begin{gather}
x = \frac{Q^2}{2p.q_\gamma} \\
y = \frac{Q^2}{sx} \\
W^2 = Q^2\frac{1-x}{x}
\end{gather}
Pseudorapidity is defined as $\eta^{*}$ = $-$ln tan $\frac{\theta}{2}$, with $\theta$ the angle between the hadron momentum and the direction of the virtual photon in the $\gamma^{*}p$ rest system.
\subsection{Proton-proton collisions at the ISR}
The second dataset used was obtained by the experiment at the CERN ISR using the SFM detector to measure momenta of all charged particles. Four samples of non-single-diffractive~(NSD) events were obtained from the $pp$ collisions at $\sqrt{s}$ = 30.4, 44.5, 52.6 and 62.2~GeV.~The experiment, data samples and event selection procedure are given in \cite{BREAKSTONE,BELL}. 

\section{\label{sec:level4} METHODOLOGY}
In the present analysis, we study the particle multiplicity distributions aiming to understand the effect of $q$-statistics. The selected data are being analyzed using the Tsallis statistics to search for the non-extensive behavior of these interactions, as described in section II. Following the details provided in the section, we devise two different methods, as described below, to solve for the nonextensivity parameter $q$ and temperature, $\beta^{-1}$.
\subsection{Method-1}
In this method $\xi$ dependence on $\beta$ is studied and the value of $\beta$ is obtained as follows:
We rearrange Eq. (\ref{eq:k}) to obtain $(q-1)$ given in Eq. (\ref{eq:qminus1}).
\begin{equation}
(q-1)=\frac{1/k+2v_0/V}{\xi^2}\label{eq:qminus1}
\end{equation}
substituting $(q-1)$ into Eq. (\ref{eq:Nbar}), we get the the following expression for $\xi$:
\begin{gather}
\xi = \frac{\alpha_1n_0}{n_0+(\alpha_1-\alpha_2)n_0^2-\langle N \rangle}  \label{eq:xi_2}\\
\mathrm{where,} \hspace{1cm}
\alpha_1=\frac{1}{k}+\frac{2v_0}{V} \label{eq:alpha_1}\\
\hspace{1.2cm}\alpha_2=\frac{2v_0}{V}    
\end{gather}
Using Eqs. (\ref{eq:n0},\ref{eq:phi}), $n_0$ can be obtained as a function of $\beta$, and putting this value of $n_0$ in Eq. (\ref{eq:xi_2}), $\xi$ can be obtained as a function of $\beta$.

The expression of $\xi$ can also be deduced from Eqs. (\ref{eq:n0},\ref{eq:phi},\ref{eq:xi}) which is based on theoretical Tsallis model and doesn't use the $\langle N \rangle$ and $k$ as used above. The steps are outlined below:
\begin{gather}
\xi = -\frac{\beta}{n_0}\frac{\partial n_0}{\partial \beta} \\
\Phi_i(\beta)  = \frac{g_im_i^2}{2\pi^2\beta}K_2(\beta m_i)
\end{gather}
where $n_0$ is given by the Eq.(\ref{eq:n0}) as below;
\begin{equation}
n_0 = V\sum_{i=1}^h \Phi_i(\beta) \\
\end{equation}
$K_n(z)$ below refers to modified bessel function of second kind. putting $n=2$ gives us an expression to calculate $\xi$ as given below : 
\begin{gather}
\frac{\partial K_n(z)}{\partial z} = -K_{n-1}(z)-\frac{n}{z}K_n(z) \\
\frac{\partial K_2(\beta m)}{\partial \beta}  = -mK_1(\beta m) -\frac{2}{\beta}K_2(\beta m) \
\end{gather}
\begin{equation}
\begin{split}
\frac{\partial\Phi_i(\beta)}{\partial\beta}=\frac{g_im_i^2}{2\pi^2}\Bigl[-\frac{K_2(\beta m_i)}{\beta^2}-\frac{1}{\beta}(m_iK_1(\beta m_i)\\
+\frac{2}{\beta}K_2(\beta m_i))\Bigr] 
\end{split}
\end{equation}
Finally we get $\xi$ as:
\begin{equation}
\xi=\frac{1}{\Sigma m_i^2g_iK_2(\beta m_i)}\Sigma m_i^2 g_i \Bigl[ 3K_2(\beta m_i)
+(\beta m_i)K_1(\beta m_i) \label{eq:xi_1} \Bigr]\ 
\end{equation}
Both these expressions of $\xi$ given by Eqs. (\ref{eq:xi_1},\ref{eq:xi_2}) are solved simultaneously by using a graphical method. Hence, plotting Eqs. (\ref{eq:xi_1},\ref{eq:xi_2}) as a function of $\beta$, the point of intersection of the two plots gives the value of ($\beta$, $\xi$). The value of $\xi$ is substituted in Eq. (\ref{eq:qminus1}) to get the value of $q$. 

\begin{figure}
\centering
\includegraphics[scale=0.18]{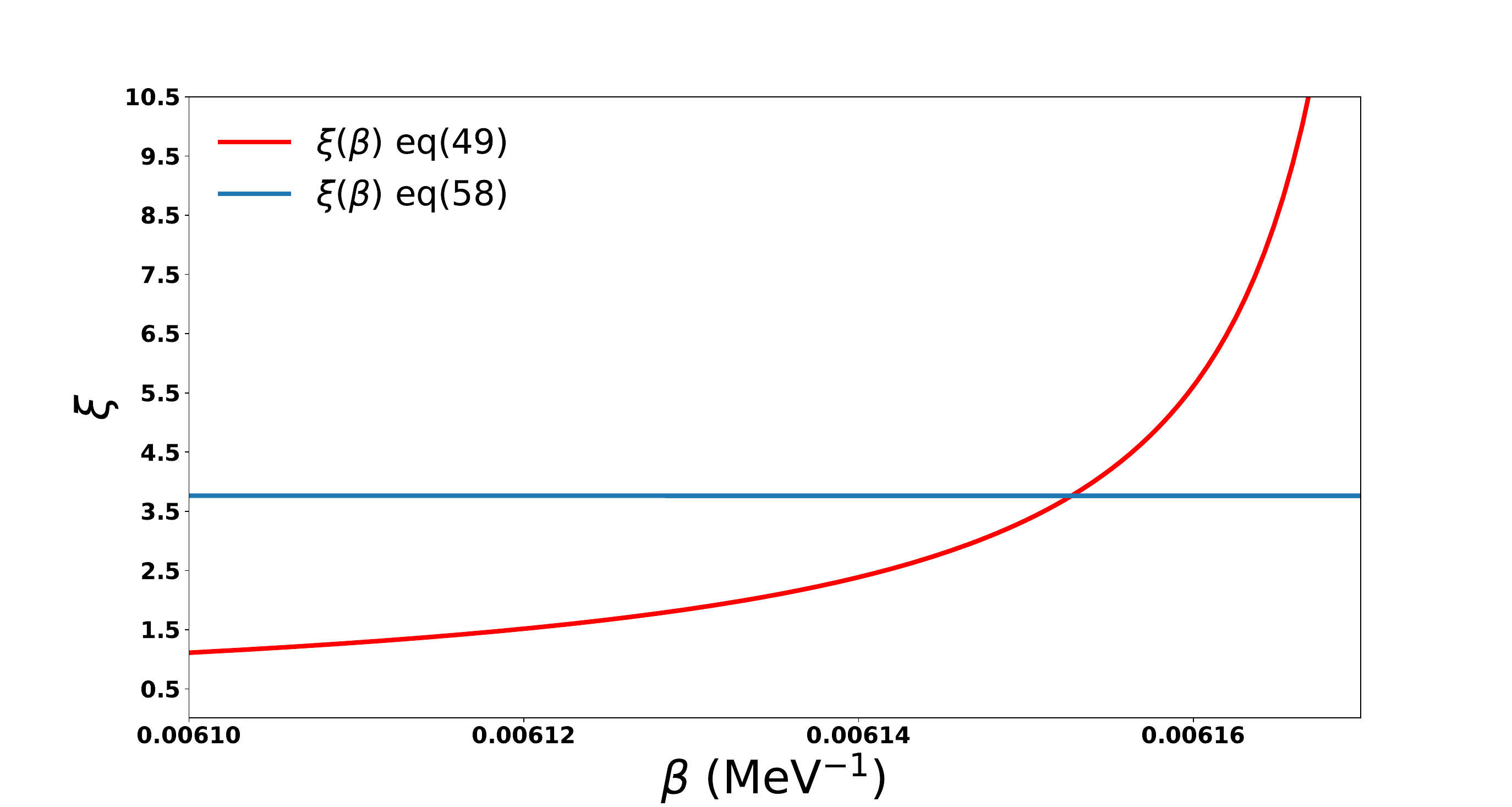}
\caption{$\xi$ versus $\beta$ for Eqs. (\ref{eq:xi_2}, \ref{eq:xi_1}) computed for $pp$ collisions at $\sqrt{s}=44.5$ GeV}
\label{fig:zb}
\end{figure}

The parameters $\langle N \rangle$ and $k$ that go into the Eqs. (\ref{eq:xi_2},\ref{eq:alpha_1}) above are obtained by fitting a negative binomial distribution to the charged multiplicity distribution. We use the data for $pp$ collisions at $\sqrt{s}$ = 44.5 GeV data from Ref. \cite{ISR1} for validation with results from C. Aguiar et. al \cite{AGUIAR2003}. The fit procedure uses CERN ROOT6.2 library. The values of $\langle N \rangle = 12.21$ and $k = 9.226$ thus obtained are given in Table \ref{tab:pp1}. Using these values, Fig. \ref{fig:zb} shows the plot of $\xi$ versus $\beta$ calculated both for Eq.~(\ref{eq:xi_2}) and for Eq.~(\ref{eq:xi_1}). Other values used for the calculations are from  Ref. \cite{AGUIAR2003} as listed below. 

\begin{itemize}
\item Excluded volume, $v_0=0.368$ fm$^3$ and Volume, $V$=40.1 fm$^3$ are used from \cite{AGUIAR2003}.
\item As described in Ref. \cite{AGUIAR2003},  $\pi^{\pm,0}$, $\eta$, $\omega$ and $\rho^{\pm,0}$ are considered as the final state hadrons,
their mass and degeneracy factor are; $m_{\pi}$ = 139.5 MeV, $m_{\eta}$ = 548.8 MeV, $m_{\rho}$ = 770 MeV and $m_{\omega}$ = 782 MeV and $g_{\pi}$ = 3, $g_{\rho}$ = 3, $g_{\eta}$ = 1 and $g_{\omega}$ = 1.  
\end{itemize}
The two curves in Fig. \ref{fig:zb} intersect at $\beta\approx$ 0.00615 MeV$^{-1}$ which gives $\beta^{-1}\approx$ 162.60 MeV. These results are in agreement with the values reported by C. Aguiar $\it{et}$ $\it{al.}$ in their paper \cite{AGUIAR2003}. The method-1 therefore validates the correctness of our procedure to calculate the entropic index $q$ and the temperature parameter, $\beta^{-1}$.
\subsection{Method-2}
In the second method, to solve for $q$ and $\beta$ we explicitly solve Eqs. (\ref{eq:Nbar}, \ref{eq:k}). Rearranging these equations, we get the following set of equations:
\begin{gather}
 n_0+(q-1)n_0\xi(n_0\xi-1)-
\frac{2v_0}{V}n_0^2-\langle N \rangle \equiv e_1\label{f1}\\
 (q-1)\xi^2-\frac{2v_0}{V}-\frac{1}{k} \equiv  e_2 \label{f2}
\end{gather}
Substituting $\xi$ from Eq.(\ref{eq:xi_1}) into Eqs.(\ref{f1}, \ref{f2}), the equations are in terms of $\beta$ and $q$. For a particular $(\beta,q)$ when $e_1 \simeq$ 0 and $e_2 \simeq$ 0, that value of $(\beta, q)$ is the desired solution. The values of parameters and constants used are the same as used in method-1. A nonlinear solver \textbf{scipy.optimize.fsolve()} (a python module Ref. \cite{2020SciPy-NMeth}) is used for solving the equations. By applying this method, we obtain the following values; \\
$e_1=-4.336 \times 10^{-12}$ and $e_2\approx-6.522 \times 10^{-16}$ \\
 $\beta\approx0.00615$ and $q\approx1.0089$\\
These values of the ($\beta,q$) pair which satisfy the equations, are consistent with the values obtained in method-1 and also with the values quoted in Ref. \cite{AGUIAR2003}. Hence it is confirmed that these methods indeed are the correct approach to solve for $\beta$  and $q$.
Having established the correctness of both methods, we use method-2 for the analysis presented in this paper. Since method-1 and method-2 give exactly the same solution for $\beta$ and $q$ values, the choice does not affect the results.
\section{\label{sec:level5}RESULTS AND DISCUSSION}
\subsection{Charged multiplicities}
\begin{figure}
\centering
\includegraphics[scale=0.19]{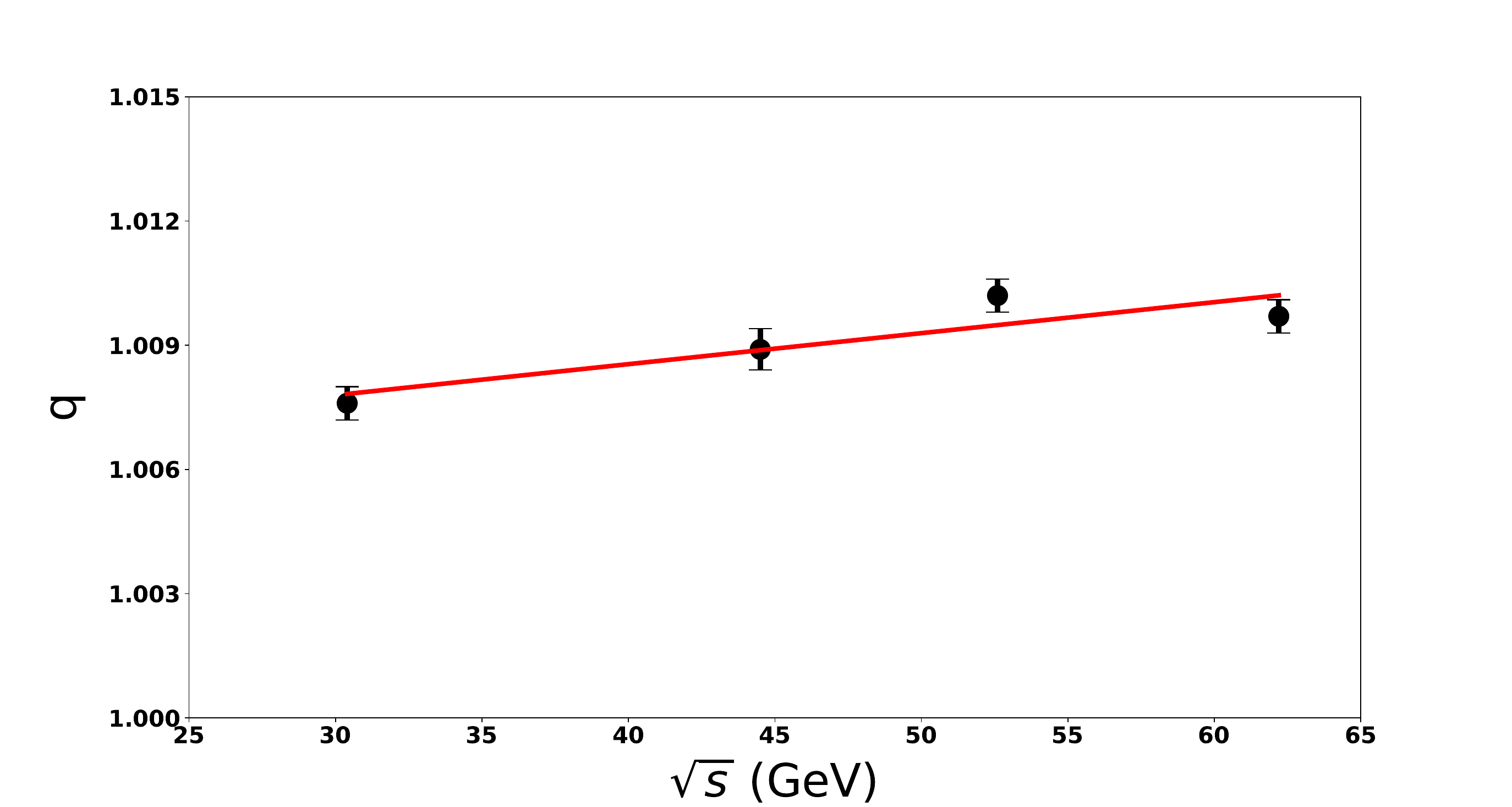}
\includegraphics[scale=0.19]{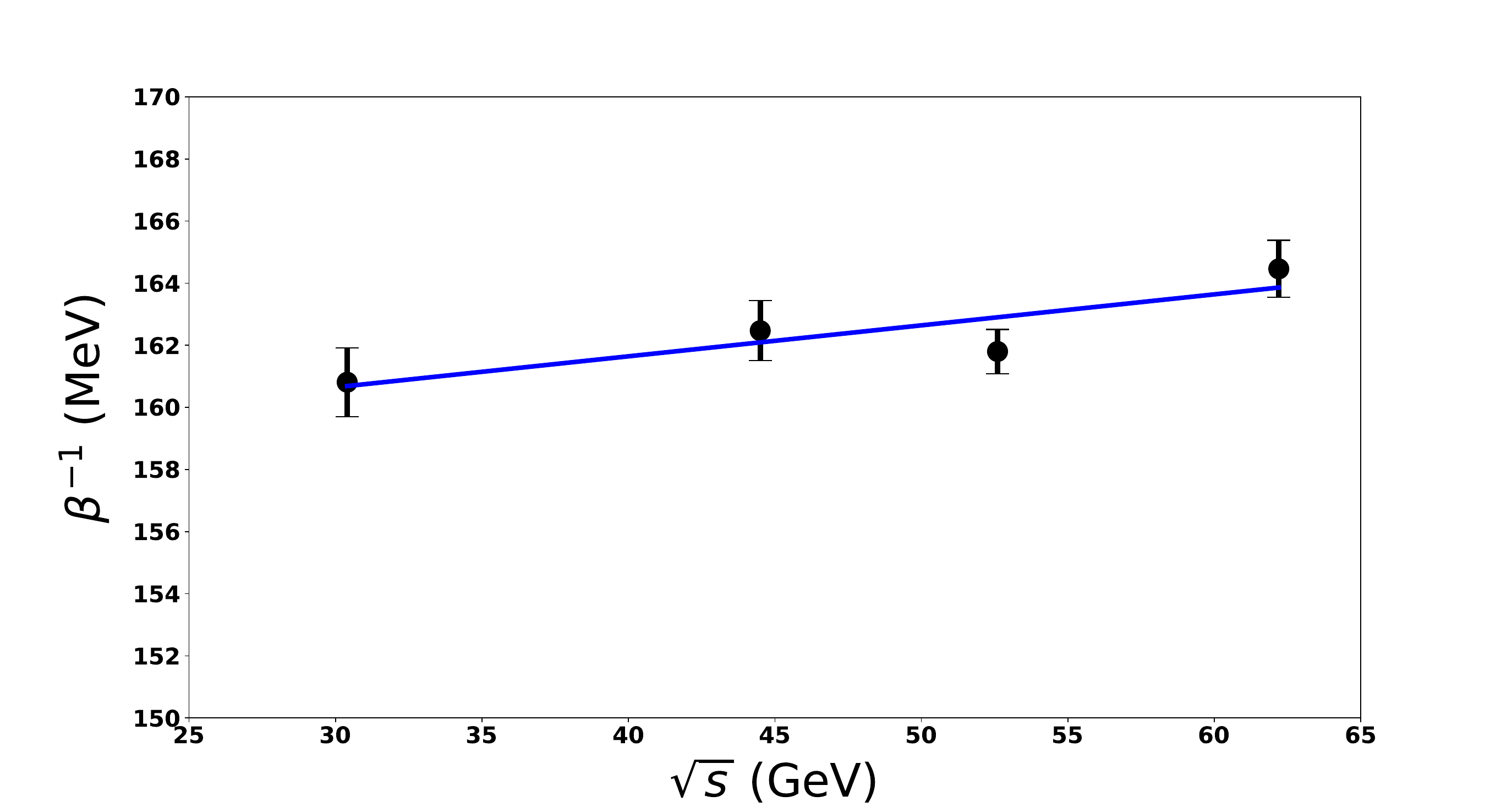}
\caption{variation of $q$ (top) and $\beta^{-1}$ (bottom) with $\sqrt{s}$ for $pp$ collisions at the ISR energies \cite{ISR1}.}
\label{fig:q_beta_pp_sqrts}
\end{figure}

\begin{figure}
\centering
\includegraphics[scale=0.4]{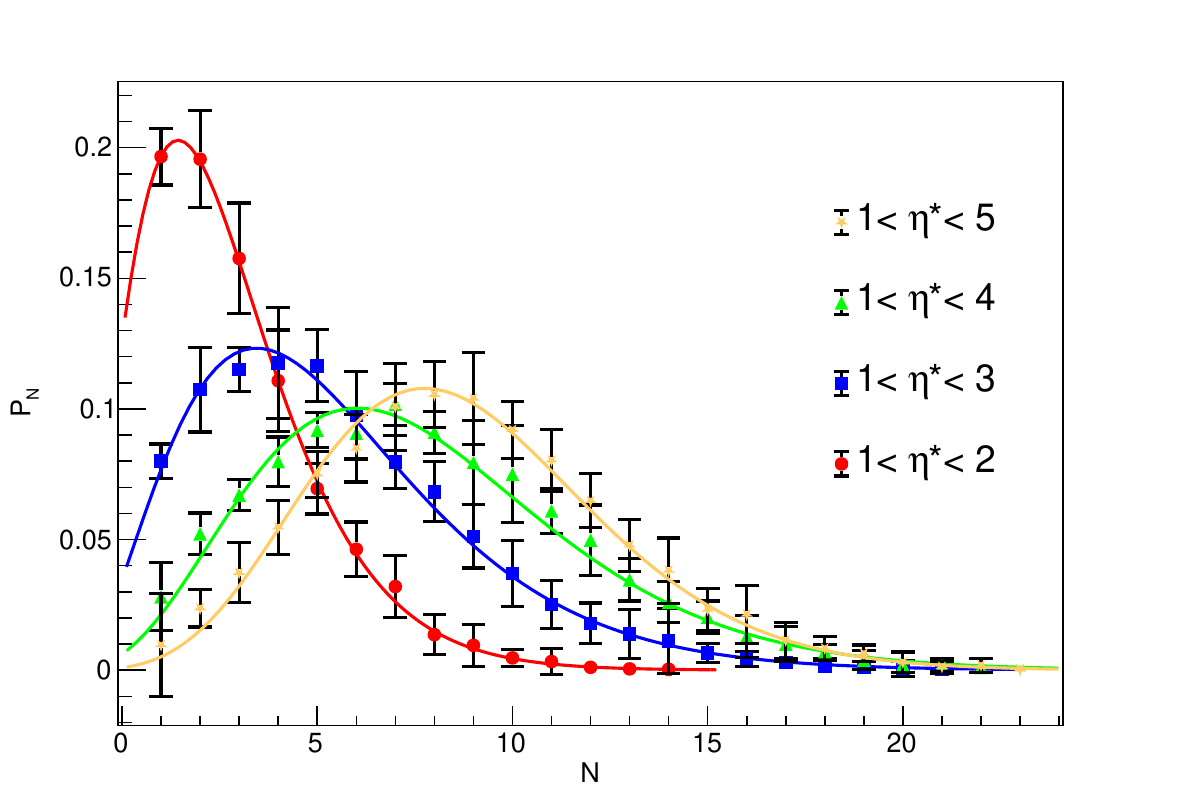}
\caption{NBD fit to the charged particle probability distribution in DIS of $ep$ for 185$<W<$220 GeV in the different pseudorapidity $\eta^*$ sectors.}
\label{fig:ept13to16Combi}
\end{figure}

As described in the two methods, the values of $q$ and $\beta^{-1}$ are validated  with the results published in \cite{AGUIAR2003} for $pp$ collisions for the 44.5 GeV data from ISR \cite{BREAKSTONE}. As further cross checks, we analyzed four sets of data from ISR range of energies by propagating experimental errors. Negative binomial distribution is fitted to the $pp$ data at $\sqrt{s}$ = 30.4, 52.6, 62.2 GeV. Table \ref{tab:pp1} gives the experimental value of average multiplicity and the NBD fit parameters. Table \ref{tab:pp2} shows the results for $q$ and $\beta^{-1}$ evaluated by using method-2. The variation of $q$ and $\beta^{-1}$ with cms energy $\sqrt{s}$ is presented in Fig. \ref{fig:q_beta_pp_sqrts}. It is observed that $q$ increases slowly and linearly with center-of-mass (cms) energy $\sqrt{s}$ as $q = a\sqrt{s} + b$  where $a=7.4849\times 10^{-5}$ and  $b=1.0055$. The parameter $\beta^{-1}$ also increases linearly with cms energy as $\beta^{-1} = c\sqrt{s}+d $ with $c$ = 0.09 and $d$= 157.65.\\
\begin{table*}
\centering
\caption{Experimental $\langle N\rangle$ and fit parameters for NBD and $\chi^2/ndf$ for $pp$ interactions at the ISR energies.}
\begin{tabular}{ c c c c c c}
\hline\hline
$\sqrt{s}$ (GeV) & $\langle N\rangle$(Expt) & $\langle N\rangle$(Fit) & $k$ & $\chi^2/ndf$  \\
\hline
30.4 & 10.54$\pm$0.14 & 10.73 $\pm$ 0.13 & 11.14 $\pm$ 0.69 & 24.04/15 \\
44.5 & 12.08$\pm$0.13 &12.21 $\pm$ 0.11 & 9.23 $\pm$ 0.49 & 11.74/17 \\
52.6 & 12.76$\pm$0.14 &12.79 $\pm$ 0.10 & 7.90 $\pm$ 0.28 &  7.93/19  \\
62.2 & 13.63 $\pm0.16$ &  13.65 $\pm$ 0.14 & 8.35 $\pm$ 0.37 & 26.32/17 \\
\hline\hline
\end{tabular}
\label{tab:pp1}
\end{table*}
It is observed from Table \ref{tab:pp2} that the variations in $q$ and $\beta^{-1}$ values on account of quoted experimental errors are very small. Hence for further analysis, the error values are not quoted.\\
\begin{table*}
\caption{$q$ and $\beta^{-1}$ values obtained from the $pp$ interactions at different ISR energies \cite{ISR1}.}
\centering
\begin{tabular}{c c c c}
\hline\hline
 $\sqrt{s}$ (GeV) & $q$ &  & $\beta^{-1}$(MeV) \\
\hline
 30.4 & 1.0076 $\pm$0.0004&   & 160.81$\pm$1.11 \\
 44.5 & 1.0089$\pm$0.0005 &   & 162.47$\pm$0.97 \\
 52.6 & 1.0102$\pm$0.0004 &   & 161.80$\pm$0.71 \\
 62.2 & 1.0097$\pm$0.0004 &   & 164.46$\pm$0.92 \\
\hline\hline
\end{tabular}
\label{tab:pp2}
\end{table*}
Charged multiplicities in $ep$ interactions in different $W$ ranges obtained by the H1 \cite{H1} experiment at HERA, are analyzed using the canonical partition function and the q-statistics. Values of entropic index $q$ and the temperature are calculated. The data in four $W$ ranges; 80-115, 115-150, 150-185 and 185-220 GeV and each range having data measured in four pseudorapidity intervals are analyzed. Thus a total of 16 datasets have been analyzed. To begin, each data is fitted with negative binomial distribution. Fig. \ref{fig:ept13to16Combi} shows the best fit results for one energy range, 185$< W<$220 GeV and four pseudorapidity sectors. Figures are given only for one invariant hadron mass range $W$ to avoid repetition of multiple similar diagrams. Table \ref{tab:H1} shows the fit results for all the data. The $\chi^2/ndf$ in Table \ref{tab:H1} shows the accuracy of fitting. $q$ and $\beta^{-1}$ values calculated by using $v_0$ = 0.368 fm$^3$ and $V$ = 40.1 fm$^3$ are shown in Table \ref{tab:H2}. 
\begin{table*} 
\caption{NBD fit parameters from Eq. (\ref{eq:nbd}) for the Charged particle multiplicity distributions in $ep$ interactions. The data are obtained by the H1 experiment~\cite{H1}.}\label{tab:H1}. 
\centering
\scalebox{0.85}{\begin{tabular}{c  c  c  c c  c   c   c }
\hline\hline
W range & $\langle W\rangle $ & $\eta^*$ &$\langle Q^2 \rangle$ & $\langle N\rangle$(Expt) & $\langle N\rangle$(Fit) & $k$ & $\chi^2/ndf$ \\
 (GeV)   & (GeV)               &     & (GeV$^2$) &   &  &   \\
\hline
80-115 & 96.9 & 1$<\eta^*<$2 & 13.9 & 2.46$\pm$0.10 & 2.44$\pm$0.11 & 3.09$\pm$0.72& 0.47/11  \\
     &   & 1$<\eta^*<$3 &27.6  & 4.90 $\pm$0.18 & 4.87$\pm$0.14 &4.71$\pm$0.63& 1.90/15  \\
 & & 1$<\eta^*<$4 & 55.0 & 6.45$\pm$0.33  &6.47$\pm$0.17 &9.78$\pm$2.02& 0.46/16  \\
 &  & 1$<\eta^*<$5 & 385.3 & 6.90$\pm$0.34  & 6.87$\pm$0.14 & 15.91$\pm$3.39 & 0.55/16  \\
\hline
115-150 & 132.0 & 1$<\eta^*<$2 & 13.9 &   2.50$\pm$0.12    & 2.54$\pm$0.07 & 3.24$\pm$0.44 & 0.60/12  \\
 &  & 1$<\eta^*<$3 & 27.5 &  5.06$\pm$0.27      & 5.10$\pm$0.14 & 3.85$\pm$0.46 & 1.79/16  \\
 &  & 1$<\eta^*<$4 & 55.1 &  7.00$\pm$0.35      & 7.07$\pm$0.13 & 7.99$\pm$0.89 & 3.51/18  \\
 &  & 1$<\eta^*<$5 & 372.8 &  7.72$\pm$0.42     &  7.68$\pm$0.12 & 15.31$\pm$2.30 & 3.75/19  \\
\hline
150-185 & 166.8 & 1$<\eta^*<$2 & 13.9 & 2.63$\pm$0.18   & 2.70$\pm$0.09 & 3.77$\pm$0.81 & 1.02/12  \\
 &  & 1$<\eta^*<$3 &  27.6 & 5.32$\pm$0.35     &  5.37$\pm$0.13 & 3.85$\pm$0.46 & 2.21/18  \\
 &  & 1$<\eta^*<$4 & 55.1 & 7.51$\pm$0.51    & 7.59$\pm$0.19 & 7.43$\pm$1.21 & 2.95/20  \\
 &  & 1$<\eta^*<$5 & 378.4 & 8.45$\pm$0.58   & 8.46$\pm$0.20 & 12.23$\pm$2.65 & 1.41/20  \\
\hline
185-220 & 201.9 & 1$<\eta^*<$2 & 13.9 & 2.66$\pm$0.18     & 2.71$\pm$0.10 & 3.62$\pm$0.78 & 0.49/12  \\
 &  & 1$<\eta^*<$3 & 27.6 & 5.35$\pm$0.35      &5.38$\pm$0.16 & 3.77$\pm$0.43 & 1.87/19  \\
 &   & 1$<\eta^*<$4 & 54.9 &7.66$\pm$0.47      & 7.76$\pm$0.16 & 6.34$\pm$0.73 &  6.00/20 \\
 &  & 1$<\eta^*<$5 & 374.2 &   8.81$\pm$0.55    & 8.83$\pm$0.18 & 15.40$\pm$2.93 & 4.22/21  \\
\hline\hline
\end{tabular}}
\end{table*}
Next we analyse both the $ep$ and $pp$ data in detail to study the relationships among various parameters. In a previous study \cite{AGUIAR2003} the $q$-statistics has been studied in $pp$ interactions by using $V$ values between 24.5-49.2 fm$^{3}$ to reproduce the experimental multiplicity distributions. It is interesting to study the variation of $\beta^{-1}$ and $q$ as a function the volume $V$.
\subsection{Variation of \texorpdfstring{$\beta^{-1}$ with $V$}{TEXT}}
\begin{figure}
\centering
\includegraphics[scale=0.20]{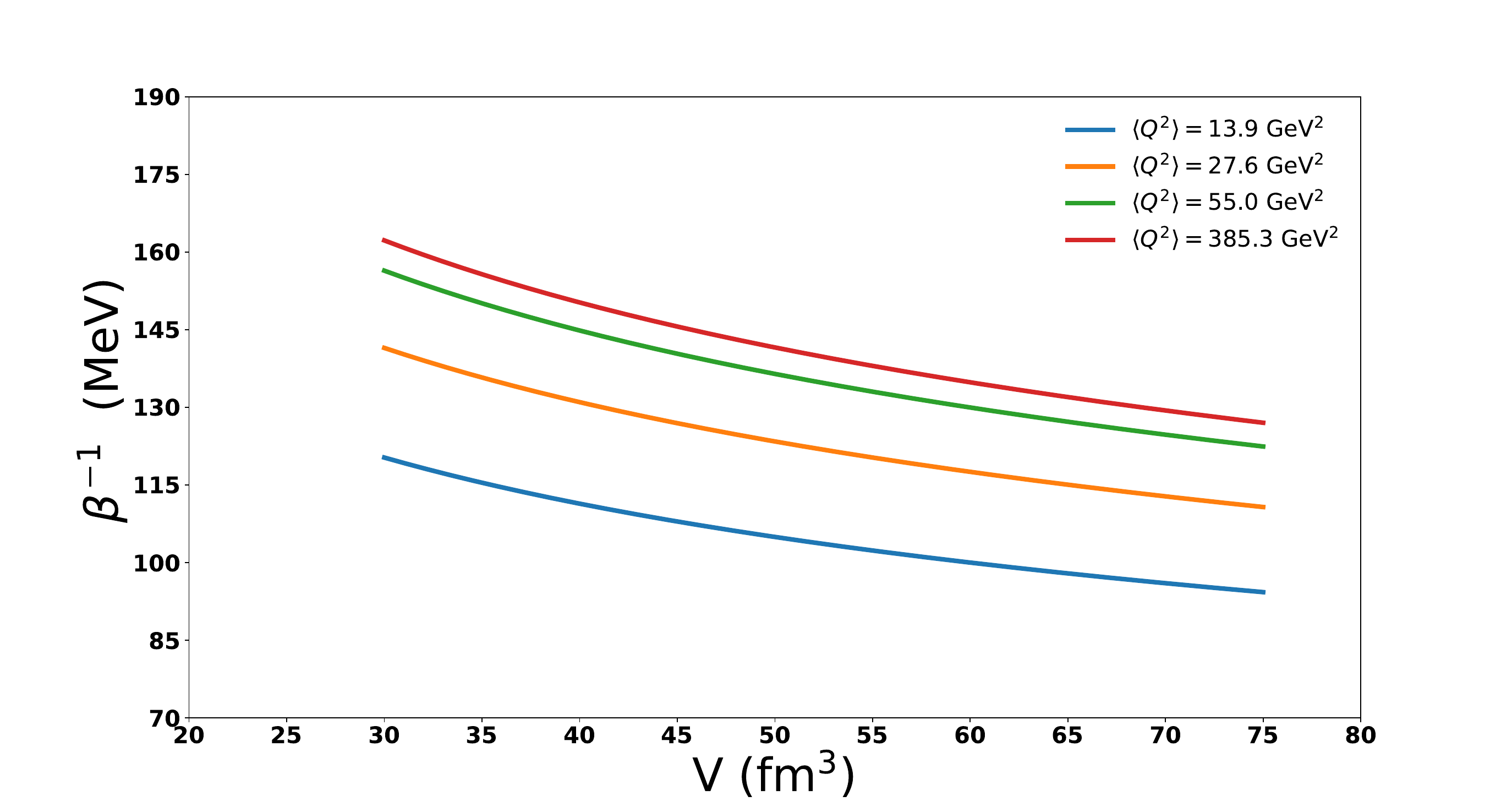}
\includegraphics[scale=0.20]{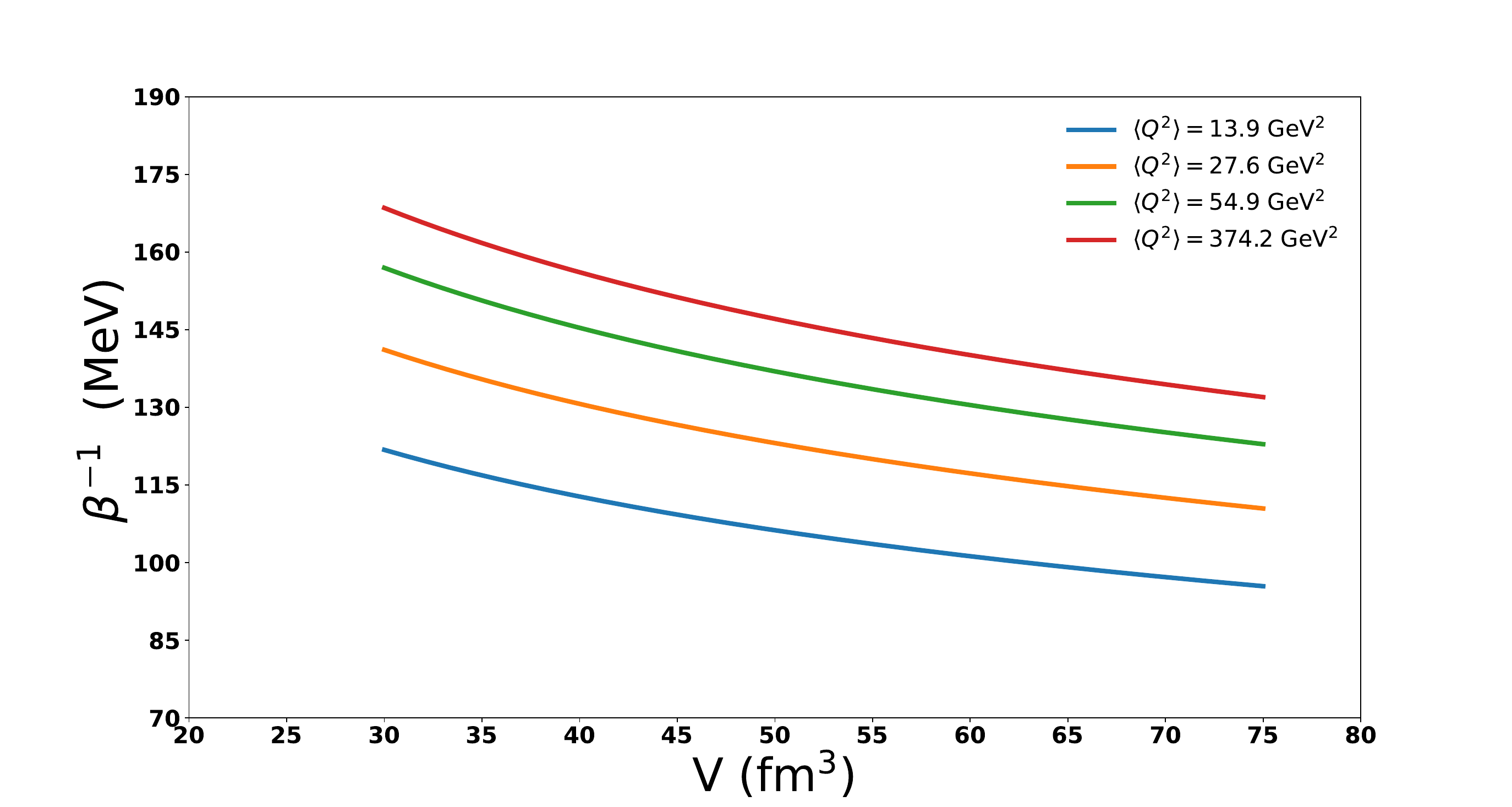}
\caption{$\beta^{-1}$ versus volume $V$ computed for $ep$ collisions in the (a) $W$= 80-115 GeV and (b) $W$= 185-220 GeV range for the four values of  
$\langle Q^{2}\rangle$.}
\label{fig:epbv}
\end{figure}

\begin{figure}
\centering
\includegraphics[scale=0.20]{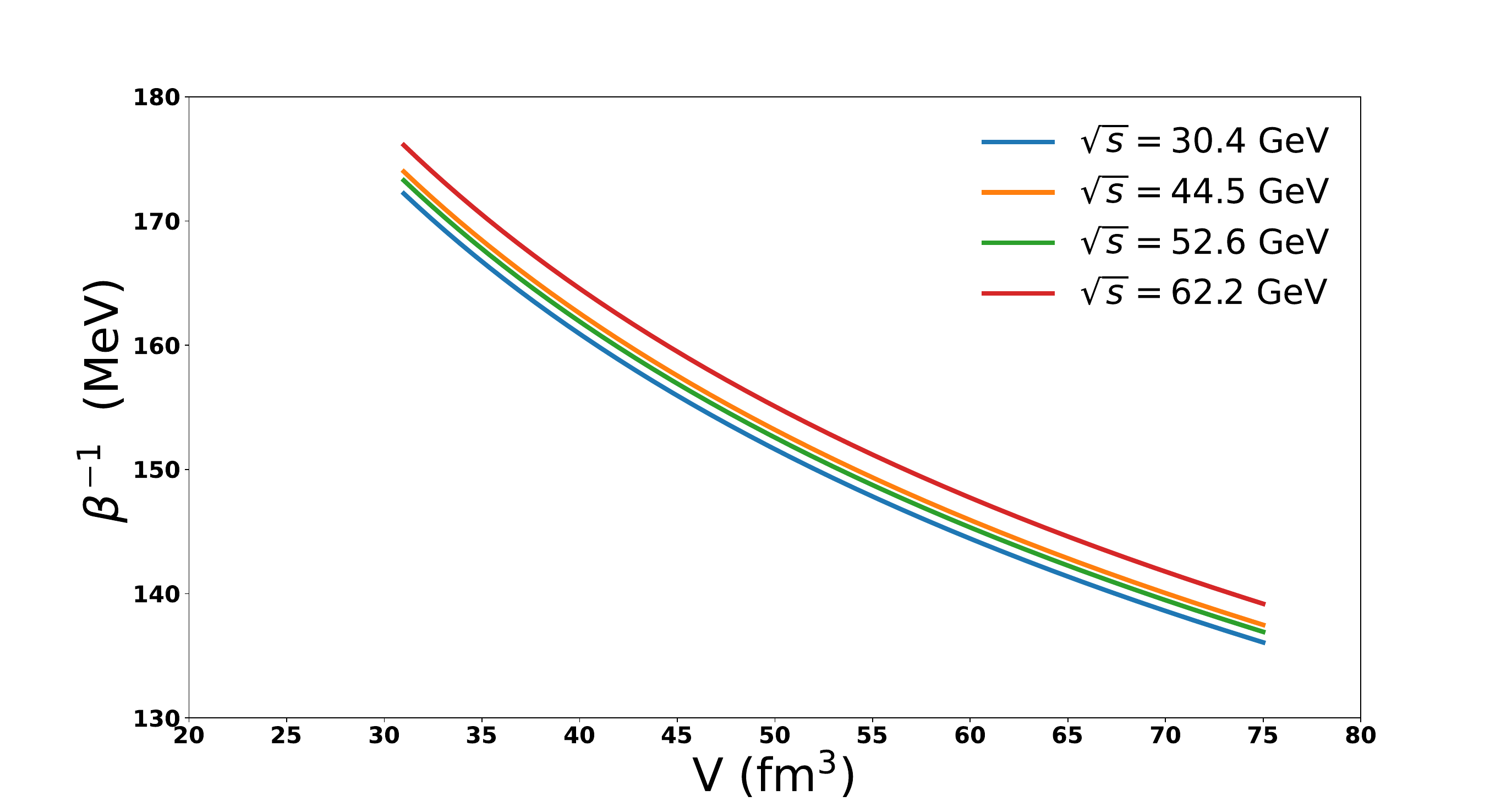}
\caption{$\beta^{-1}$ versus volume $V$ computed for $pp$ collisions at four $\sqrt{s}$=30.5, 44.5, 52.6, 62.2 GeV.}
\label{fig:ppbv}
\end{figure}

Fig. \ref{fig:epbv} shows a plot of $\beta^{-1}$ as a function of volume $V$ for $ep$ collisions in one pseudorapidity sector, $1<\eta^{*}<2$ for W  = 80-115, 185-220 GeV ranges, corresponding to the lowest and the highest $\langle W\rangle$ values. Similar distributions are studied for all $W$-ranges and of $\eta^{*}$-pseudorapidity sectors. Figures are not included to omit repetition. It is observed that the temperature($\beta^{-1}$) decreases slowly with the increase in the volume for all ranges of $W$. We also studied the variation of $\beta^{-1}$ for the different values of volume $V$ = 30-75 fm$^{3}$ in four pseudorapidity sectors for the lowest and the highest $\langle W\rangle$ values. It is again observed that, $\beta^{-1}$ decreases with volume $V$. With the increase in volume of the system, the temperature decreases by $\simeq$ 25\%  as the volume changes from 30 to 75 fm$^{3}$ for every $\langle Q^{2}\rangle$. However, at a given value of $V$ the temperature increases with $\langle Q^{2}\rangle$. 
Fig. \ref{fig:ppbv} shows the dependence $\beta^{-1}$ on V for $pp$ collisions at different cms energies. It is observed that for $pp$ interactions, $\beta^{-1}$ falls with volume $V$ i.e with the increase in volume of the system, the temperature decreases for each cms energy. However, at a fixed value of the volume, the temperature rises with $\sqrt{s}$. From the $p_T$ spectrum study using Tsallis function, the results from the CMS experiment \cite{s39} also show that $T (\beta^{-1})$ increases with $\sqrt{s}$. Thus the temperature dependence on volume is very similar in both type of interactions, $ep$ and $pp$. 
\begin{table}[h]
\caption{ Values of $q$ and $\beta^{-1}$ for the $ep$ interactions from the data obtained by the H1 \cite{H1} experiment in different $W$ ranges.}\label{tab:H2}
\centering
\scalebox{0.85}{\begin{tabular}{c  c  c  c   c   c   }
\hline\hline
W range   & $\langle W\rangle$  & $\eta^*$ & $\langle Q^2 \rangle$   &  $q$   & $\beta^{-1}$  \\
 (GeV)   & (GeV)               &     & (GeV$^2$) &     &  (MeV) \\
\hline
80-115 & 96.9 & 1$<\eta^*<$2 & 13.9  & 1.0228  & 111.37 \\
       &   & 1$<\eta^*<$3    &27.6  & 1.0152  & 130.92 \\
 & & 1$<\eta^*<$4 & 55.0  & 1.0080 & 144.76 \\
 &  & 1$<\eta^*<$5 & 385.3  & 1.0057 &  150.17 \\
\hline
115-150 & 132.0 & 1$<\eta^*<$2 & 13.9 & 1.0243 &  111.22 \\
 &  & 1$<\eta^*<$3 & 27.5 & 1.0187 & 129.64 \\
 &  & 1$<\eta^*<$4 & 55.1 & 1.0096 &  145.28 \\
 &  & 1$<\eta^*<$5 & 372.8 & 1.0063 &  152.90 \\
\hline
150-185 & 166.8 & 1$<\eta^*<$2 & 13.9 & 1.0243  & 112.34 \\
 &  & 1$<\eta^*<$3 &  27.6 & 1.0197 &  130.22 \\
 &  & 1$<\eta^*<$4 & 55.1 & 1.0112 &  145.45 \\
 &  & 1$<\eta^*<$5 & 378.4 & 1.0069 &  154.52 \\
\hline
185-220 & 201.9 & 1$<\eta^*<$2 & 13.9 & 1.0239 & 112.70 \\
 &  & 1$<\eta^*<$3 & 27.6 & 1.0193  & 130.59 \\
 &   & 1$<\eta^*<$4 & 54.9 & 1.0119  & 145.27 \\
 &  & 1$<\eta^*<$5 & 374.2 & 1.0068  & 156.01 \\
\hline\hline
\end{tabular}}
\end{table}
\subsection{Variation of \texorpdfstring{$q$ with $V$}{TEXT}}
\begin{figure}
\centering
\includegraphics[scale=0.20]{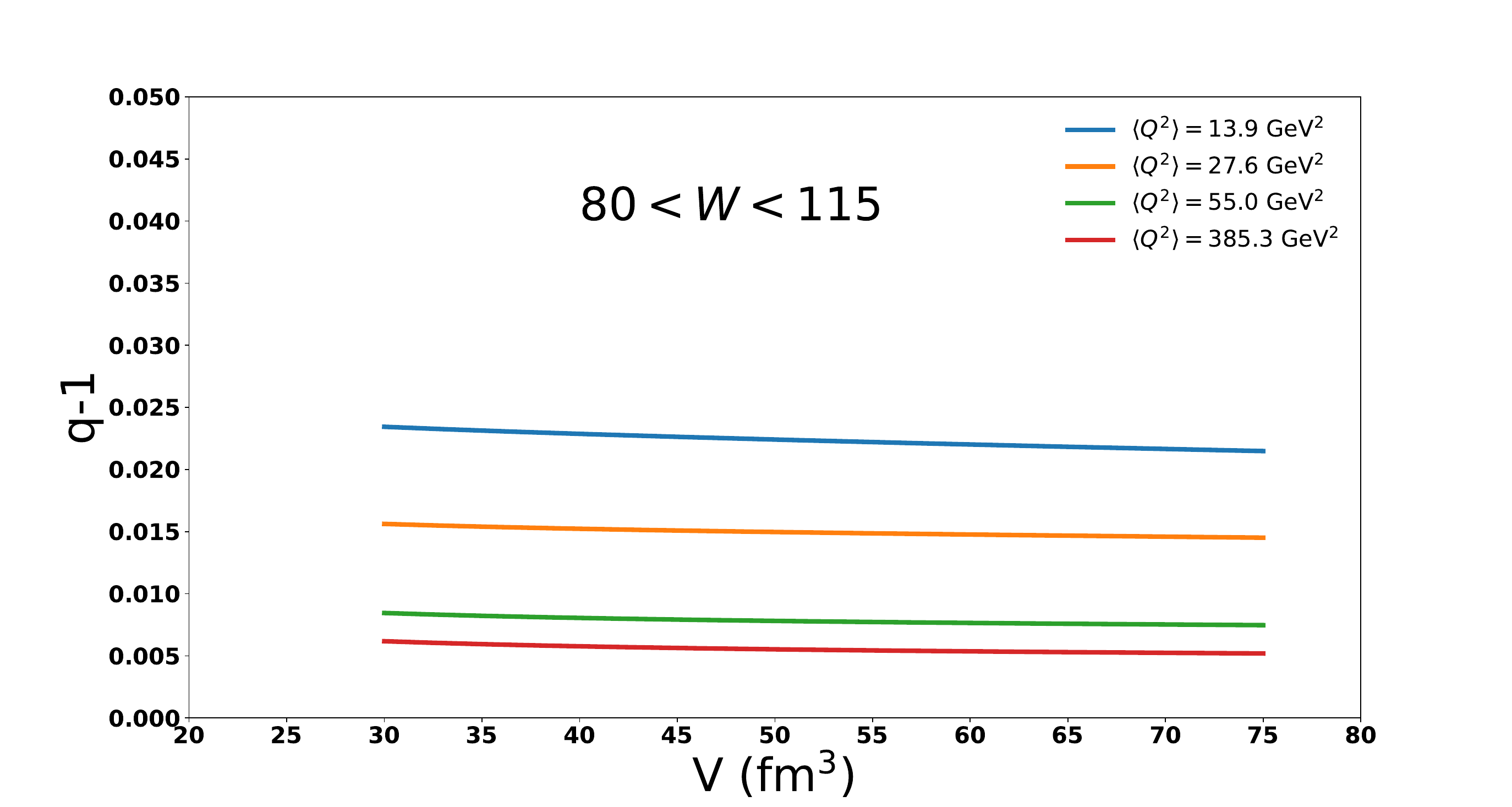}
\includegraphics[scale=0.20]{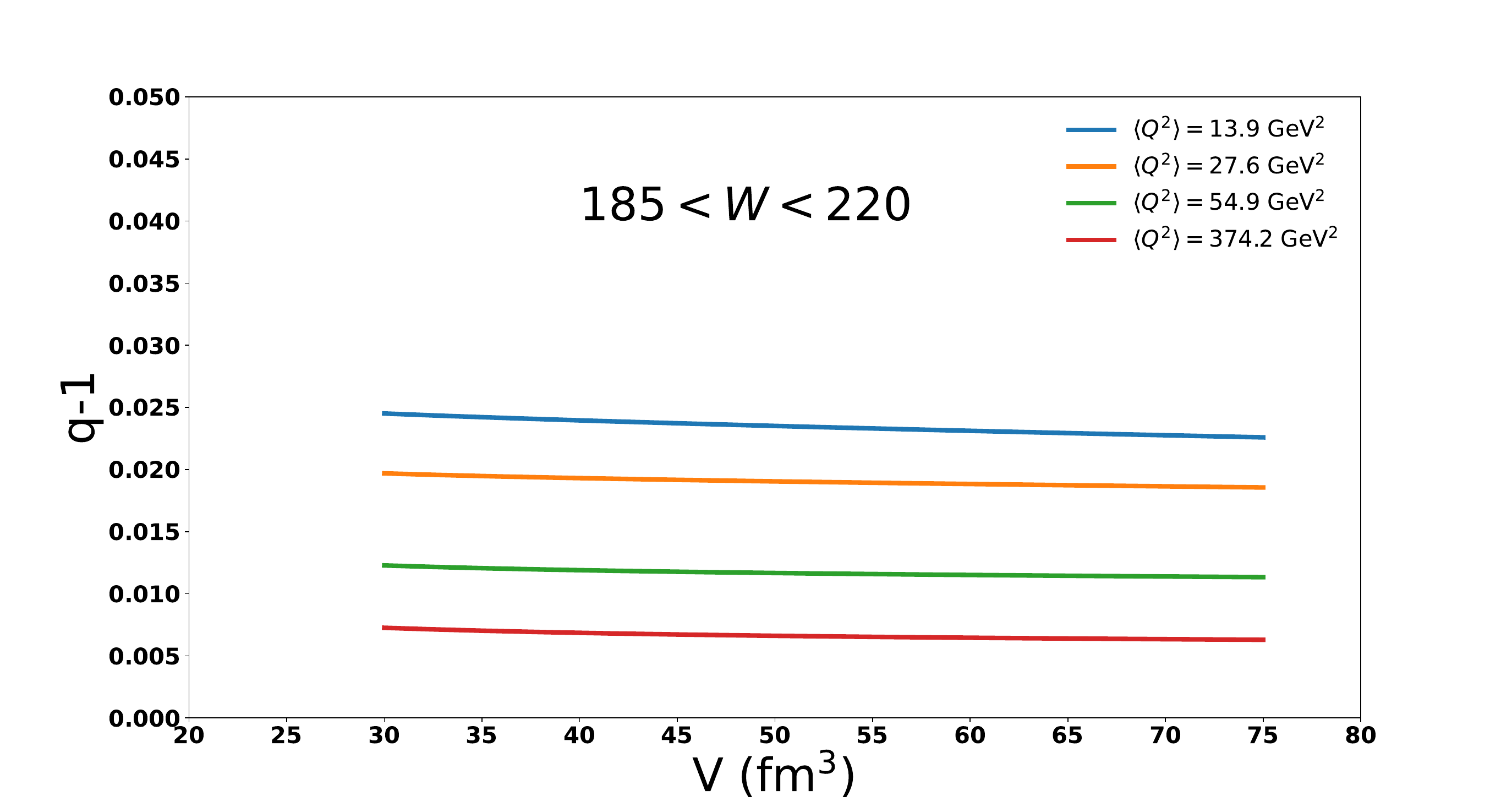}
\caption{$(q-1)$ versus $V$ dependence for $ep$ collisions in (a) $W$ = 80-115 GeV and (b) $W$ = 185-220 GeV range and with different $\langle Q^{2}\rangle$.}
\label{fig:t1epqV}
\end{figure}

 Fig. \ref{fig:t1epqV} shows the variation of $(q-1)$ with volume $V$ for $ep$ collisions in $W$ = 80-115 GeV and $W$ = 185-220 GeV range and with different $\langle Q^{2}\rangle$ values having pseudorapidity in $1<\eta^{*}<2$ interval. It is observed that $q$ has nearly no dependence on $V$ for each $\langle Q^{2}\rangle$ value in the mentioned pseudorapidity region. Same trend is found in all pseudorapidity regions and $W$ ranges. The $(q-1)$ decreases marginally with the increase in $\langle Q^{2}\rangle$ for every $V$. However $(q-1) >0$ is always true which shows canonical entropy characteristically nonextensive.

\begin{figure}
\centering
\includegraphics[scale=0.20]{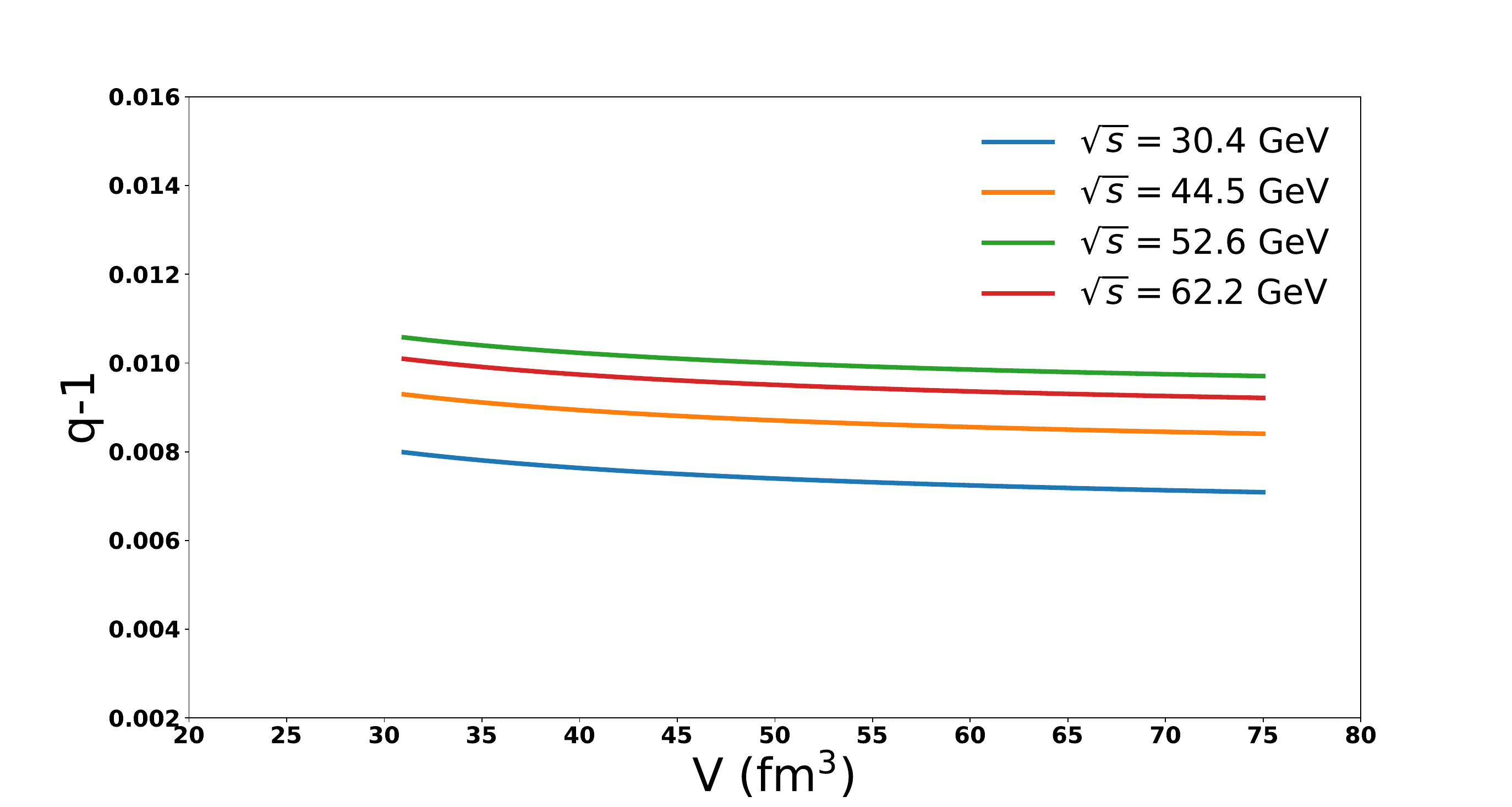}
\caption{Variation of $(q-1)$ with $V$ for $pp$ collisions at different energies}
\label{fig:ppqV}
\end{figure}

Fig. \ref{fig:ppqV} shows the ($q-1$) variation with $V$ for $pp$ collisions at different cms energies. Similar to the case of $ep$ interactions, $(q-1)$ for each $\sqrt{s}$ is nearly constant for all $V$. Similar trend persists for all cms energies. However $q$ is always greater than unity, at every $V$ and every $\sqrt{s}$. The $(q-1)$ value increases with the cms energy $\sqrt{s}$. Thus, as the cms energy of interaction increases, the nonextensive behavior of the system becomes more predominant.

\subsection{Variation of \texorpdfstring{$q$ and $\beta^{-1}$ with $\langle W\rangle$}{TEXT}}
\begin{figure}
    \centering
    \includegraphics[scale=0.20]{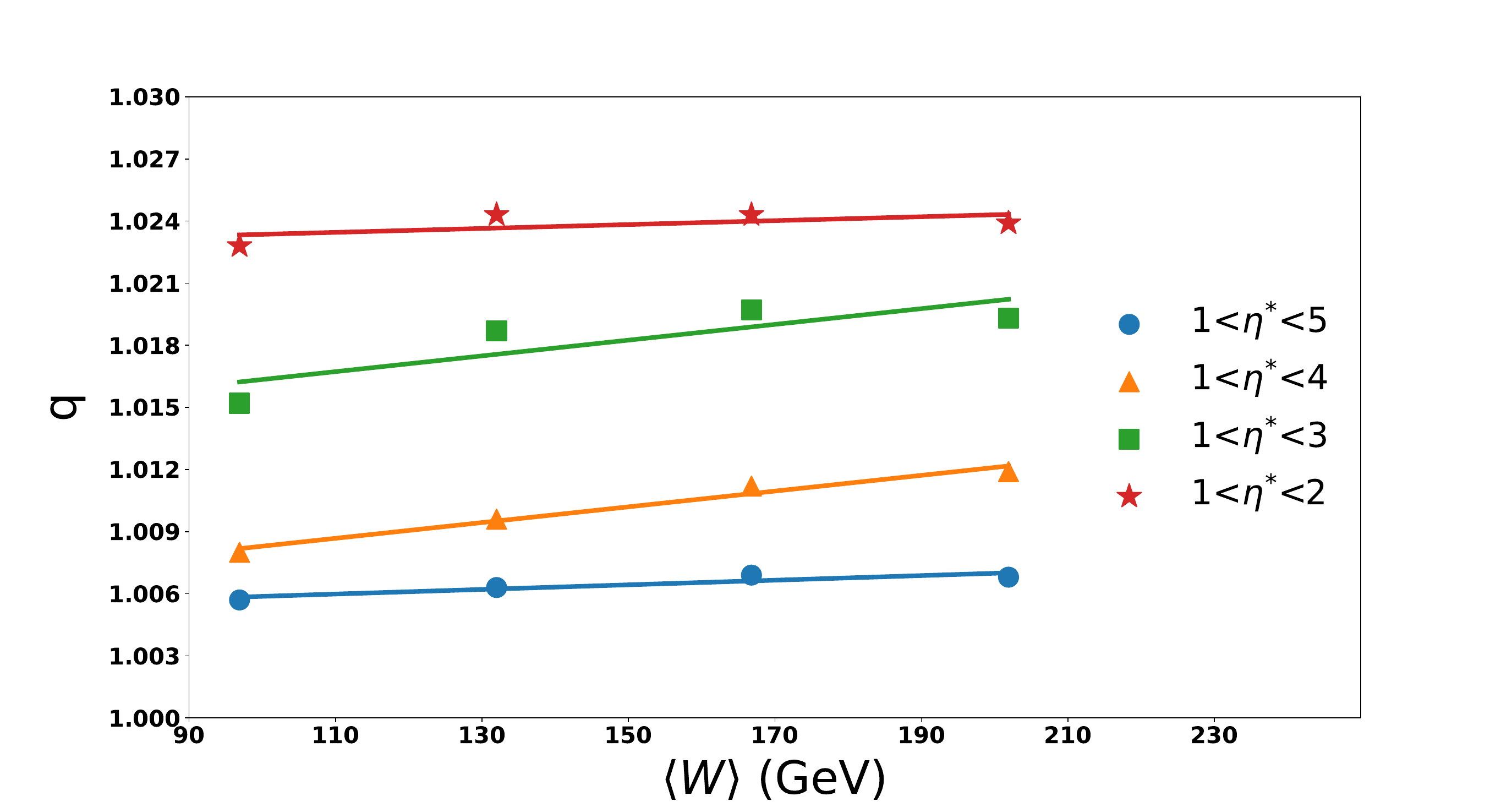}
    \includegraphics[scale=0.20]{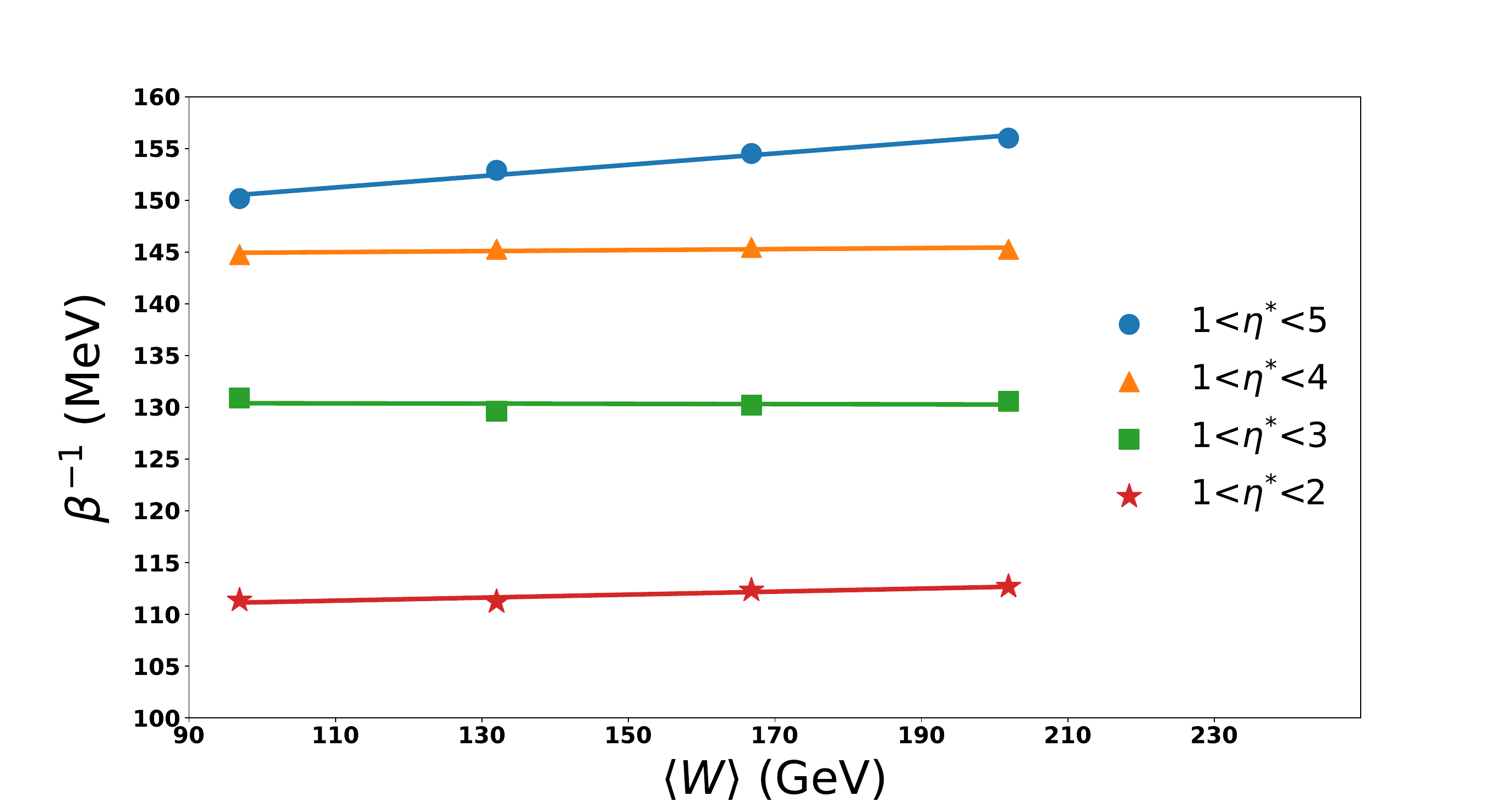}
    \caption{Variations of $q$ and $\beta^{-1}$ versus $\langle W\rangle$ for $ep$ interactions. $\langle W\rangle$= 96.9, 132.0, 166.8, 201.9 GeV corresponding to the  W range 80-115, 115-150, 150-185, 185-220 GeV respectively.}
    \label{fig:q_ep_beta_W}
\end{figure}

To study the dependence of the entropic index $q$ on the hadronic mass $W$, we show in Fig. \ref{fig:q_ep_beta_W} the variation plot for $ep$ interactions in different $\eta^{*}$ sectors. It is observed that $q$ varies slightly but almost linearly with $\langle W\rangle$ in each phase space region shown by the pseudorapidity range. The fit parameters of the linear fit are given in Table \ref{tab:qVsW}.\\
\begin{table} 
\caption{Linear variation plot $q = a\langle W\rangle +b$ for $ep$ interactions in different $\eta^{*}$ sectors, from Table \ref{tab:H2}. \hspace{1cm}\\} \label{tab:qVsW} 
\centering
\begin{tabular}{c  c  c }\\
\hline\hline
$\eta^*$ & $a$ ($\times 10^{-5})$ & $b$ \\
\hline
$1<\eta^{*}<5$  & 1.115 & 1.0047\\
$1<\eta^{*}<4$  & 3.802 & 1.0044\\
$1<\eta^{*}<3$  & 3.803 & 1.0125\\
$1<\eta^{*}<2$  & 0.939 & 1.0224\\
\hline\hline
\end{tabular}
\end{table}
$\beta^{-1}$ also depends linearly on $\langle W\rangle$ with a marginal change. However both $q$ and $\beta^{-1}$ depend upon the phase space size. While $q$ decreases, $\beta^{-1}$ increases with increasing $\eta^{*}$ region. The linear fit is parameters are given in Table \ref{tab:betaVsW}.
\begin{table} 
\caption{Linear variation plot $\beta^{-1}=a\langle W\rangle$+b for $ep$ interactions in different $\eta^{*}$ sectors, from Table \ref{tab:H2}. \hspace{1cm}\\}\label{tab:betaVsW}
\centering
\begin{tabular}{c  c  c }\\
\hline\hline
$\eta^{*}$ & $a$ ($\times 10^{-4})$ & $b$\\
\hline
$1<\eta^{*}<5$  & 547 & 145.225\\
$1<\eta^{*}<4$  &  48 & 144.464\\
$1<\eta^{*}<3$  & -12 & 130.519 \\
$1<\eta^{*}<2$  & 145& 109.727\\
\hline\hline
\end{tabular}
\end{table}
\subsection{Variation of \texorpdfstring{$q$ and $\beta^{-1}$ with $\langle Q^{2}\rangle$}{TEXT}}
\begin{figure}
\centering
\includegraphics[scale=0.20]{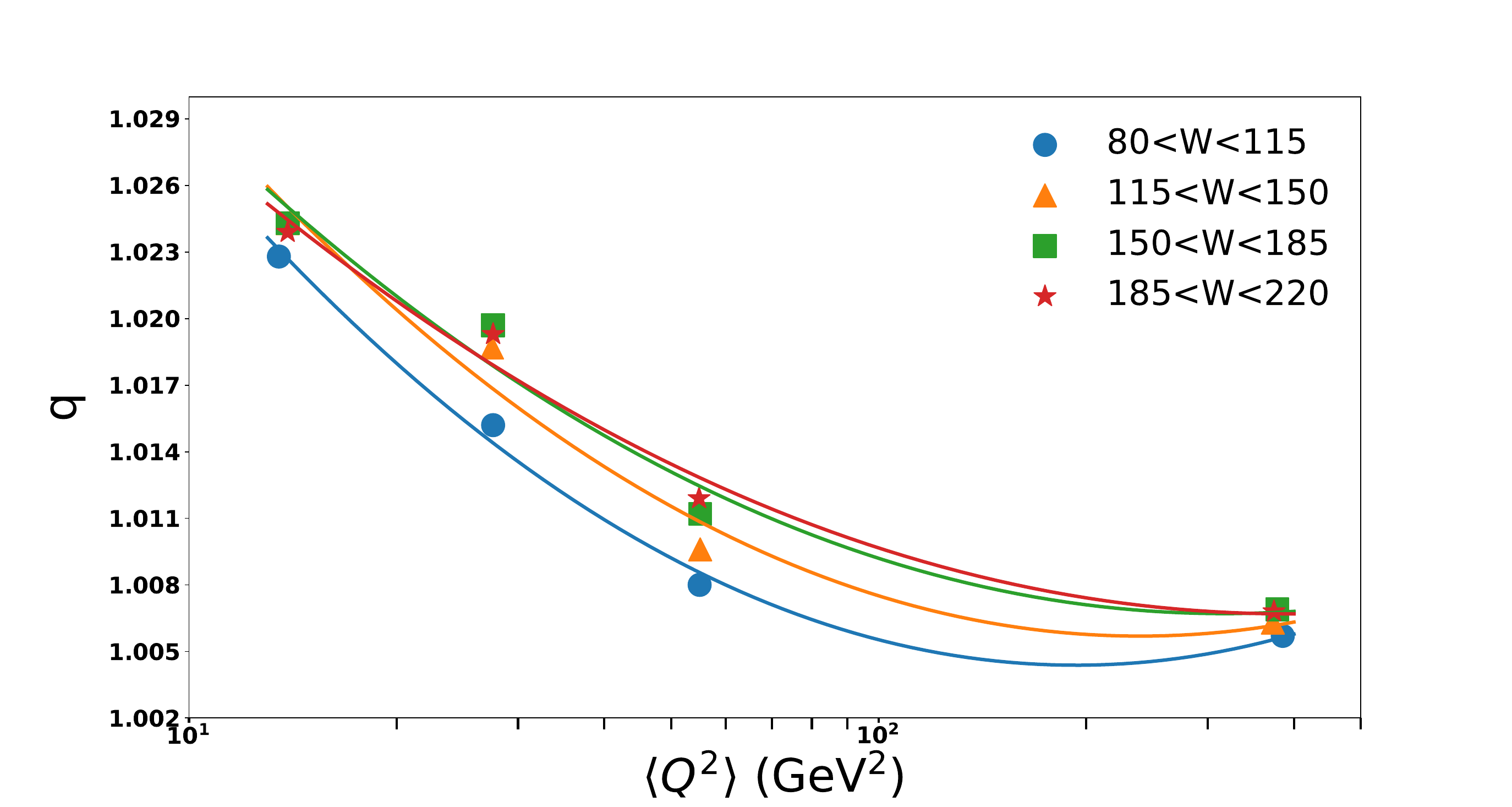}
\includegraphics[scale=0.20]{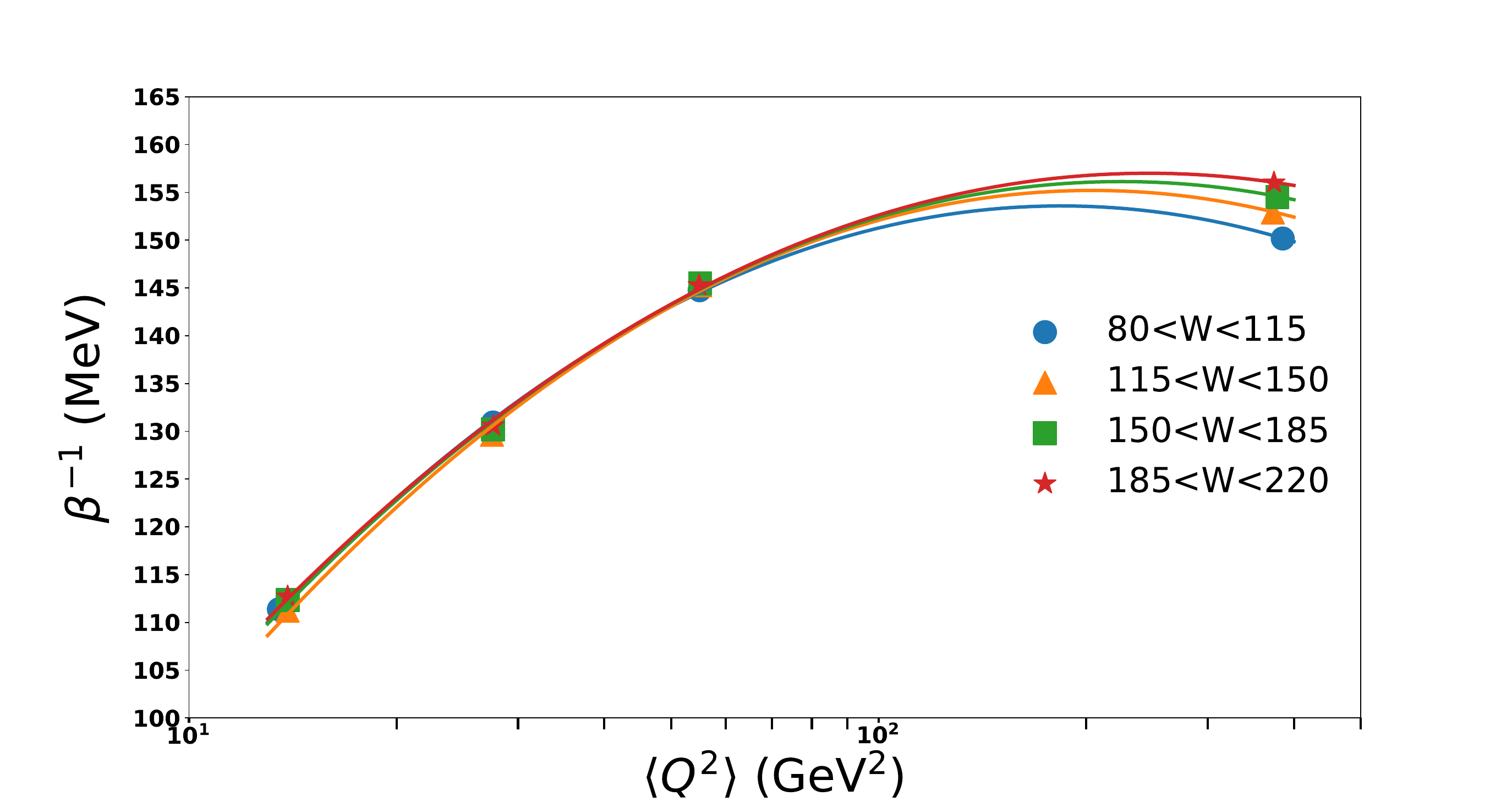}
\caption{$q$ and $\beta^{-1}$ versus $\langle Q^2 \rangle$ variation for all $ep$ interactions in different $\langle W\rangle$ values listed in the Table \ref{tab:H2}}
\label{fig:ep_q_beta_Q2}
\end{figure}

The variations of $q$ and $\beta^{-1}$ as a function of $\langle Q^{2}\rangle$, the negative square of the four-momentum transferred from the electron to the proton, are shown in Fig. \ref{fig:ep_q_beta_Q2}. The variation of $q$ has a quadratic dependence on log$\langle Q^{2}\rangle$ with parameters given in Table \ref{tab:qQuad}.\\
\begin{table} 
\caption{Variation plot $q=a\log^{2}\langle Q^2 \rangle-b\log\langle Q^2 \rangle+c$ for $ep$ interactions in different $W$ ranges. \hspace{1cm}\\}\label{tab:qQuad} 
\centering
\begin{tabular}{c  c  c c}\\
\hline\hline
$W$(GeV) & $a$ ($\times 10^{-4})$ & $b$($\times 10^{-4})$ &$c$\\
\hline
$ 80<W<115$  & 140 & 641 & 1.0776\\
$115<W<150$  & 126 & 602 & 1.0773\\
$150<W<185$  & 99  & 497 & 1.0688\\
$185<W<220$  & 84  & 438 & 1.0634\\
\hline\hline
\end{tabular}
\end{table}
\begin{table} 
\caption{ plot $\beta^{-1}=-a^{\prime}\log^{2}\langle Q^2 \rangle+ b^{\prime}\log\langle Q^2 \rangle-c^{\prime}$ for $ep$ interactions in different $W$ ranges \hspace{1cm}\\}\label{tab:betaQuad} 
\centering
\begin{tabular}{c  c  c c}\\
\hline\hline
$W$(GeV) & $a^{\prime}$  & $b^{\prime}$ & $c^{\prime}$\\
\hline
$ 80<W<115$  & 32.89 & 148.99 & 15.14\\
$115<W<150$  & 32.58 & 150.52 & 18.61\\
$150<W<185$  & 30.21 & 142.09 & 10.94\\
$185<W<220$  & 28.59 & 136.73 & 06.47\\
\hline\hline
\end{tabular}
\end{table}
It may be observed that the $q$ decreases quadratically with $\log\langle Q^{2}\rangle$. For a given $\langle Q^{2}\rangle$ the $q$ value is higher for the interactions from higher $W$ range i.e higher $\langle W\rangle$. However for all values of $\langle Q^{2}\rangle$ the $q$ remains greater than unity ($>1$). This indicates the nonextensive dynamics of interactions.

On the other hand, temperature $\beta^{-1}$ rises quadratically with log$\langle Q^{2}\rangle$. The parameters of fit are given in Table \ref{tab:betaQuad}. Also at the highest $\langle Q^{2}\rangle$ value, the temperature is the highest for maximum $\langle W\rangle$. For a given $\langle W\rangle$ four-momentum transferred from the electron to the proton in an $ep$ interaction increases with the phase space size i.e from $|\Delta\eta^{*}|$ = 1 to 4. Correspondingly the increase in the energy in the region results in increase in temperature. 
\section{\label{sec:level6} CONCLUSION}
The paper presents the first analysis of $ep$ interactions at $\sqrt{s}$=300 GeV and $pp$ interactions at various ISR energies by using the canonical partition function and $q$-statistics. We devised two different methods for finding the entropic parameter $q$ and the temperature parameter $\beta^{-1}$. The two methods are validated with the previously published results \cite{AGUIAR2003} and are found to be consistent. The analysis of $ep$ interactions in different ranges of the invariant mass of the hadronic system shows that the interactions deviate from Maxwell-Boltzmann statistics and show that the canonical entropy is nonextensive. The entropic parameter $q >1$ for all $W$ ranges and in all pseudorapidity sectors between $\eta^{*}$=1 to 5. Similarly the analysis of $pp$ interactions establishes that entropic parameter $q >1$ for all cms energies. From a detailed study of the interdependence of $q$, $\beta$, $V$, $\sqrt{s}$, $\langle W\rangle$ and $\langle Q^{2}\rangle$, we observe that for $ep$ interactions,  with the increase in volume of the system, the temperature decreases at a given $\langle Q^{2}\rangle$. However, for fixed volume $V$, the temperature rises with  $\langle Q^{2}\rangle$. The temperature dependence on volume is very similar in $ep$ and $pp$ interactions as observed from Figs. \ref{fig:epbv},\ref{fig:ppbv}. It is observed that $q$ has nearly no dependence on system volume for both $ep$ and $pp$ interactions. However $q>1$ is always true, which shows that the canonical entropy is characteristically nonextensive. 

In $ep$ interactions, $q$ has a linear dependence on $\langle W\rangle$ and the value of $q$ decreases with the increase in allowed phase space i.e. as the size of $\eta^{*}$ sector increases. The temperature parameter $\beta^{-1}$ depends marginally upon hadronic invariant mass $\langle W\rangle$ but $\beta^{-1}$ increases with the size of the phase space, roughly from 111 MeV to 155 MeV as the $\eta^{*}$ changes from $1<\eta^*<2$ to $1<\eta^*<5$. For a given $\langle W\rangle$, square of the four-momentum transferred from the electron to the proton $\langle Q^{2}\rangle$ in an $ep$ interaction increases with the phase space size ($|\Delta\eta^{*}|$) as can be observed from Table \ref{tab:H2}. Correspondingly the temperature ($\beta^{-1}$) also increases. In case of $pp$ interactions $q$ has linear dependence on $\sqrt{s}$, similar to the $ep$ case. The temperature increases with the cms energy, $\sqrt{s}$. The results from the CMS experiment \cite{s39,s40} have also shown that the temperature increases with the  energy, $\sqrt{s}$.

It is concluded that canonical entropy derived in both $ep$ and $pp$ interactions is nonextensive, with entropic parameter $q>$1. In the Van der Waals gas model with $q$=1, the multiplicity distribution is narrower in width and does not agree with the data \cite{AGUIAR2003}. The Tsallis $q$-statistics broadens the width and reproduces the multiplicity distribution well. This also implies an increase in the events with  higher multiplicity. In the limit of small $(q-1)$, the distributions are well represented by negative binomial distribution. The present work has a potential scope of extension to the $ep$ and electron-nucleus collisions at the future Electron-Ion Collider(EIC) and to $pp$ and proton-nucleus collisions at the High Luminosity (HL)-LHC experiments. The present study is validated using the  $pp$ collisions data at ISR energies and the entropic parameter is found to increase slightly with the energy, $\sqrt{s}$. The HL-LHC experiments will facilitate to explore the nature of entropic parameter and study the temperature parameter, $\beta^{-1}$, at higher $\sqrt{s}$ energies. The entropic parameter based on present study can be predicted to be non-extensive for $pp$ collisions at higher energies. The future planned experiments at the EIC will enable to explore the nonextensive nature of entropic parameter, and temperature $\beta^{-1}$ in $ep$ collisions, at lower values of hadronic invariant mass $\langle W\rangle$ as compared to the present study. 
\section{ACKNOWLEDGMENT} 
Author Soumya Sarkar acknowledges the support from the Department of Science and Technology, Govt. of India for the DST Inspire fellowship.
\bibliography{TS.bib}
\end{document}